\documentclass{ametsocV6.1_nolinenumbers}
\usepackage{svg}
\usepackage{stfloats}
\usepackage{array}
\usepackage{lipsum} 
\usepackage{graphicx}
\usepackage{subcaption}
\usepackage{caption}
\usepackage{booktabs}
\usepackage{comment}
\usepackage{placeins}

\begin{document} 

\title{Exploring coupled tropical Pacific variability within a Multi-branch $\beta$-Variational Autoencoder}

\authors{Emily F. Wisinski,\aff{a}\correspondingauthor{Emily Wisinski, ewisinsk@umd.edu} 
Maria J. Molina,\aff{a} 
Kyle J. C. Hall,\aff{a} 
Hannah Bao,\aff{a} 
Salil Mahajan,\aff{b}
Nan Rosenbloom,\aff{c}
and John Fasullo\aff{c} 
}

\affiliation{
\aff{a}{Department of Atmospheric and Oceanic Science, University of Maryland, College Park, MD, USA}\\
\aff{b}{Oak Ridge National Laboratory, Oak Ridge, TN, USA}\\
\aff{c}{U.S. NSF National Center for Atmospheric Research, Boulder, CO, USA}\\
}



\abstract{This study explores what is encoded in the latent space of a multi-branch $\beta$-variational autoencoder ($\beta$-VAE) trained on coupled tropical Pacific climate fields. We assess the reconstruction skill and physical interpretability of the latent space trained on monthly sea surface temperature, ocean heat content, and outgoing longwave radiation across the tropical Pacific from a 500-year preindustrial control simulation. The model generalizes well, with only modest degradation from training to test performance, and preserves the dominant basin-scale structure of all three fields. Latent-space diagnostics show that variability is organized unevenly across dimensions: sea surface temperature is concentrated in a smaller subset of latent dimensions, whereas ocean heat content and outgoing longwave radiation are more broadly distributed across multiple dimensions. Comparisons with conventional tropical Pacific diagnostics further show that several latent dimensions align with known El Niño and La Niña variability, while others capture related coupled ocean-atmosphere variability on decadal or longer timescales. Sensitivity experiments and latent traversals identify dimensions associated with eastern-Pacific-like, central-Pacific-like, coastal, subsurface-dominant, and atmosphere-dominant variability. Together, these results show that the multi-branch $\beta$-variational autoencoder yields a skillful and physically informative reduced representation of coupled tropical Pacific variability.}


\maketitle
\statement
This study asks whether artificial intelligence can learn a compact and physically meaningful representation of climate variability in the tropical Pacific. We trained a model on sea surface temperature, upper-ocean heat content, and cloud-related radiation, then assessed whether its learned patterns aligned with known tropical Pacific behavior. The model organized these fields into a small set of patterns linked to different parts of the ocean–atmosphere system. Several patterns were closely tied to El Niño and La Niña, while others captured related variability on longer timescales or were not captured by standard indices alone. These results show that artificial intelligence can compress climate information into interpretable patterns and may support future studies of climate variability and predictability.

%
\section{Introduction}

A variational autoencoder (VAE) is an unsupervised probabilistic framework for nonlinear dimensionality reduction that learns a low-dimensional representation of high-dimensional data while supporting reconstruction of the original input data \citep{Goodfellow-et-al-2016,han2025climate}. Its generative capability arises from combining an encoder-decoder architecture with a latent prior, which enables sampling and interpolation in the latent space. A $\beta$-variational autoencoder ($\beta$-VAE) modifies the standard VAE objective by scaling the Kullback–Leibler divergence term with a scalar $\beta$, thereby adjusting the strength of regularization toward the prior \citep{higgins2017betavae}. In the Earth systems, VAEs and related architectures are increasingly used for dimensionality reduction, emulation, and generative modeling \citep[e.g.,][]{doshi2025unsupervised,ma2025modeling,han2025climate,king2025leveraging,macmillan2025towards}. However, many applications emphasize compression or prediction, while giving less attention to the physical structure encoded in the learned latent representations and to whether that structure can be interpreted in terms of known climate processes \citep[e.g.,][]{paccal2025understanding}. This interpretive gap is particularly relevant in the tropical Pacific, where climate variability arises from strongly coupled interactions among the ocean subsurface, sea surface, and overlying atmosphere across timescales.

Variability in the tropical Pacific spans multiple spatiotemporal scales, reflecting both slowly evolving oceanic processes and more rapidly varying atmospheric behavior \citep{deser2010sea}. A reduced nonlinear representation that jointly captures these coupled fields could be useful not only for data compression and subsequent reconstruction, but also for diagnosing how variability is organized in a latent space and whether that organization is physically meaningful. 

In this context, the El Niño-Southern Oscillation (ENSO) provides a natural benchmark because it is the dominant mode of tropical Pacific variability and involves coupled surface, subsurface, and atmospheric anomalies. ENSO is commonly described in terms of warm (El Niño) and cool (La Niña) phases that recur on interannual timescales, typically every 2 to 7 years \citep{mcphaden2006enso}. However, ENSO is not a fixed pattern. Decades of research have underscored the rich diversity of ENSO events, with an emphasis on differences in the zonal extent of sea surface temperature anomalies (SSTAs) and variations in associated atmospheric and subsurface responses \citep[e.g.,][]{rasmusson1982variations,ashok2007nino,kao2009contrasting,capotondi2015understanding,pan2025diversity}. More broadly, ENSO diversity reflects nonlinear coupled dynamics, event asymmetries, and interactions across the ocean–atmosphere system that are not fully described by any single index or linear decomposition \citep{takahashi2011enso,dommenget2013analysis,timmermann2018nino,williams2018diversity,geng2019atmospheric,srinivas2024dominant}.

As a result, many methods have been used to characterize the structure and diversity of ENSO. Common approaches rely on surface temperature-based regional indices (e.g., Niño-3 and Niño-4) or pattern-based diagnostics and principal component methods that isolate dominant spatial variability \citep{kug2009two,yeh2009nino,lemmon2019metric,kao2009contrasting,yu2012changing,takahashi2011enso}. Other studies have expanded the analysis to variables such as subsurface ocean temperature \citep{yu2011subsurface}, sea surface salinity \citep{singh2011contrasting,qu2014enso}, and outgoing longwave radiation \citep{chiodi2013nino}. These approaches have produced important insights, but many are fundamentally linear or rely on limited geographic domains, making it difficult to represent nonlinear relationships, multivariate coupling, or variability distributed across multiple fields. This challenge motivates the use of machine learning methods that can learn nonlinear structure directly from data while retaining the possibility of physical interpretation \citep{reichstein2019deep,molina2023review}. Recent advances have applied deep neural networks and explainable AI for a wide variety of ENSO applications, including forecasting, identifying precursors, and revealing ENSO dynamics, yet evaluating physical interpretability within unsupervised generative frameworks remains underexplored \citep{ham2019deep,toms2020physically,shin2022application,deng2024explainable,chen2025toward,chen2025combined}.

Here we apply a multivariate $\beta$-VAE to fields spanning the ocean subsurface, sea surface, and overlying atmosphere, with a separate branch for each variable (i.e., multi-branch), thereby sampling key components of coupled tropical Pacific variability. Our goal is not only to obtain a reduced representation, but also to determine whether the learned latent space is physically interpretable. Accordingly, our objectives are to (i) evaluate the reconstruction skill of a multi-branch $\beta$-VAE applied to the tropical Pacific; (ii) diagnose how information is organized across the learned latent space; and (iii) interpret the latent dimensions in relation to known tropical Pacific-related patterns, timescales, and ocean-atmosphere processes.

\section{Data and Methods}

\subsection{Earth System Model Data}

Since the advent of the satellite era ($\approx$1979), only about 14 El Niño and 14 La Niña events have been observed, making observational data alone insufficient for training a $\beta$-VAE aimed to describe the diversity of ENSO \citep{furtado2026setting}. Divergent observational and model responses to external forcing \citep[e.g.,][]{watanabe2021enhanced} complicate established approaches to addressing sample-size issues, such as transfer learning \citep{ham2019deep}. To mitigate both sample-size constraints and uncertainty surrounding the ENSO response to external forcing, we adopt a `perfect model framework,' designating a 500-year preindustrial control simulation (piControl) as the `ground truth' for $\beta$-VAE training.

We use the Energy Exascale Earth System Model version 2 \citep[E3SMv2;][]{golaz2022doe} from the U.S. Department of Energy, which captures key aspects of Pacific variability across timescales, although biases remain in the amplitude and timing of simulated variability \citep{fasullo2024modes,hall2025knowledge}. Relative to E3SMv1, these biases are generally comparable or reduced, and the model shows improved representation of clouds, precipitation, and ENSO-related processes \citep{golaz2022doe,fasullo2024modes}. The 500-year E3SMv2 piControl also exhibits substantially reduced drift relative to its predecessor, though some spatially varying long-term drift remains \citep{fasullo2023overview}. The atmospheric component uses a nominal 1-degree grid with 72 vertical levels, while the ocean component employs a variable-resolution approach, with a coarser grid (60km) at midlatitudes and a finer grid (30 km) in the equatorial and polar regions \citep{fasullo2023overview}. 

\subsection{Preprocessing Data}

\begin{figure*}[t]
    \centering
    \includegraphics[width=\textwidth]{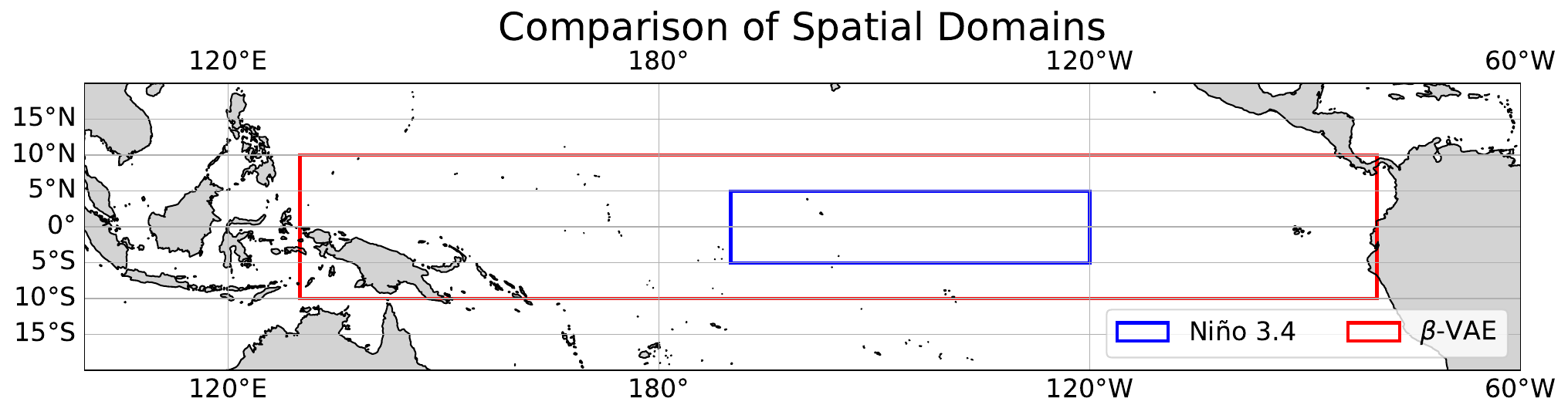}
    \caption{Comparison of the spatial domain of the Niño-3.4 region (blue) and the $\beta$-VAE (red).}
    \label{fig:domains}
\end{figure*}

Monthly atmospheric and oceanic variables from the E3SMv2 piControl were selected as input features for the $\beta$-VAE, including outgoing longwave radiation (OLR), which serves as a proxy for deep convection, along with sea surface temperatures (SSTs) and ocean heat content (OHC; 0-700m). OHC is generated at monthly intervals from the native E3SM ocean model grid using a constant specific heat ($c$) of 3990 $J\,kg^{-1}\,C^{-1}$ and an assumed density ($\rho$) of 1026 $kg\,m^{-3}$. Fields are re-gridded to a 1-degree grid using conservative interpolation. Variables were subset to the region of interest (10$^{\circ}$N-10$^{\circ}$S, 130$^{\circ}$E-80$^{\circ}$W; Fig. \ref{fig:domains}), which spans the Maritime Continent to Peru, to allow the $\beta$-VAE to capture variability across the Pacific basin contrary to the more constrained geographic bounds used for ENSO monitoring (Niño-3.4). Within the chosen domain, ENSO is expected to be the dominant mode of variability captured in the $\beta$-VAE latent space. The region is constrained to the tropical Pacific, and extending beyond may require input variables beyond SST, OHC, and OLR to adequately capture additional modes of climate variability. Similarly, further filtering of the data (e.g., a high-pass filter to isolate lower-frequency modes) can be applied to limit the influence of dominant modes, such as ENSO, but also imposes additional data preprocessing steps that may limit the $\beta$-VAE's ability to learn relationships in the latent space.

Any missing or invalid data were removed, primarily from land grid cells in SSTs and OHC, resulting in variable dimensions that differ (Table \ref{tab:vae}). A monthly climatology for the 500-year simulation was computed for each variable, and anomalies were calculated by subtracting the monthly climatology from the monthly data. Detrending was not necessary because piControls have no external forcing and therefore exhibit no global long-term trends. Each month is treated as a single `sample' for the $\beta$-VAE.

A regional trend was identified for OHC \citep[potentially related to upper-ocean model drift;][]{fasullo2023overview}, which could bias the patterns learned by the $\beta$-VAE if fed recursively. Since the regional trend was isolated to OHC, we did not apply detrending because doing so would require variable-specific treatment and additional assumptions about the trend’s origin. Thus, the monthly anomalies were randomly split into a training set (70\%; 4800 samples) and a test set (30\%; 1200 samples), rather than by temporal blocks, to minimize preprocessing and reduce learned autocorrelation biases. While it is common in predictive settings to compute anomalies using statistics derived from the training set only \citep{furtado2026setting}, our goal is not out-of-sample prediction but rather the latent-space characterization of tropical Pacific variability, where any dependence between the training and test sets is negligible for our latent-space interpretations. A z-score variant was used to standardize the data, thereby handling outliers: 

\begin{equation}
X_{\text{scale}} = \frac{X_{i} - X_{\text{med}}}{X_{75} - X_{25}}, 
\end{equation} 

\noindent where $X_i$ is the training set sample, $X_{med}$ is the training set median, and $X_{25}$ and $X_{75}$ are the training set 25th and 75th percentiles (i.e., interquartile range). The test set was standardized using the training set's median and percentiles.

\begin{table*}[t]
\centering
\caption{$\beta$-VAE hyperparameters. The validation set is a percentage of the training set.}
\begin{tabular}{l c c}
Hyperparameter & $\beta$-VAE \\
\hline
SST Vector Length & 2968 \\
OHC Vector Length & 2960 \\
OLR Vector Length & 2662 \\
Number of Encoder Hidden Layers & 5 \\
Number of Decoder Hidden Layers & 5 \\
Number of Encoder Nodes per Layer & 400; 250; 180; 100; 30\\
Number of Decoder Nodes per Layer & 30; 100; 180; 250; 400\\
Number of Latent Space Nodes & 20 \\
Activation Functions & Hyperbolic Tangent \\
KL Divergence Weight ($\beta$) & 0.0005 \\
Loss Function & MSE + KL Divergence \citep{odaibo2019tutorial} \\
Optimizer & Adam \citep{kingma2014adam} \\
Learning Rate & 0.0003 \\
Training Epochs & 150 \\
Mini-batch Size & 256 \\
Validation Size & 25\% \\
\end{tabular}
\label{tab:vae}
\end{table*}

The Oceanic Niño Index (ONI) was computed for the E3SMv2 piControl to benchmark the $\beta$-VAE latent space against a known representation of ENSO. ONI was the primary index used operationally by the National Oceanic and Atmospheric Administration (NOAA) to monitor ENSO \citep{bamston1997documentation} and remains an essential benchmark in climate-related research. To compute ONI, SSTs were subset to the Niño-3.4 region (5$^{\circ}$N-5$^{\circ}$S, 170$^{\circ}$W-120$^{\circ}$W; Fig. \ref{fig:domains}). Spatially averaged SSTAs were then computed relative to the 500-year E3SMv2 monthly climatology; a 30-year climatology updated every 5 years, previously used by NOAA for ONI, is not needed due to piControl stationarity. A 3-month centered rolling mean was applied to the SSTAs to reduce higher-frequency variability. Events were designated as El Niño when a $+0.5^{\circ}$C threshold was reached or exceeded for five consecutive months. Events were similarly designated as La Niña using a $-0.5^{\circ}$C threshold. Events that did not meet either criterion were labeled as ENSO neutral.

\subsection{Multi-branch $\beta$-VAE Architecture}

\begin{figure*}[t]
    \centering
    \includegraphics[width=\textwidth]{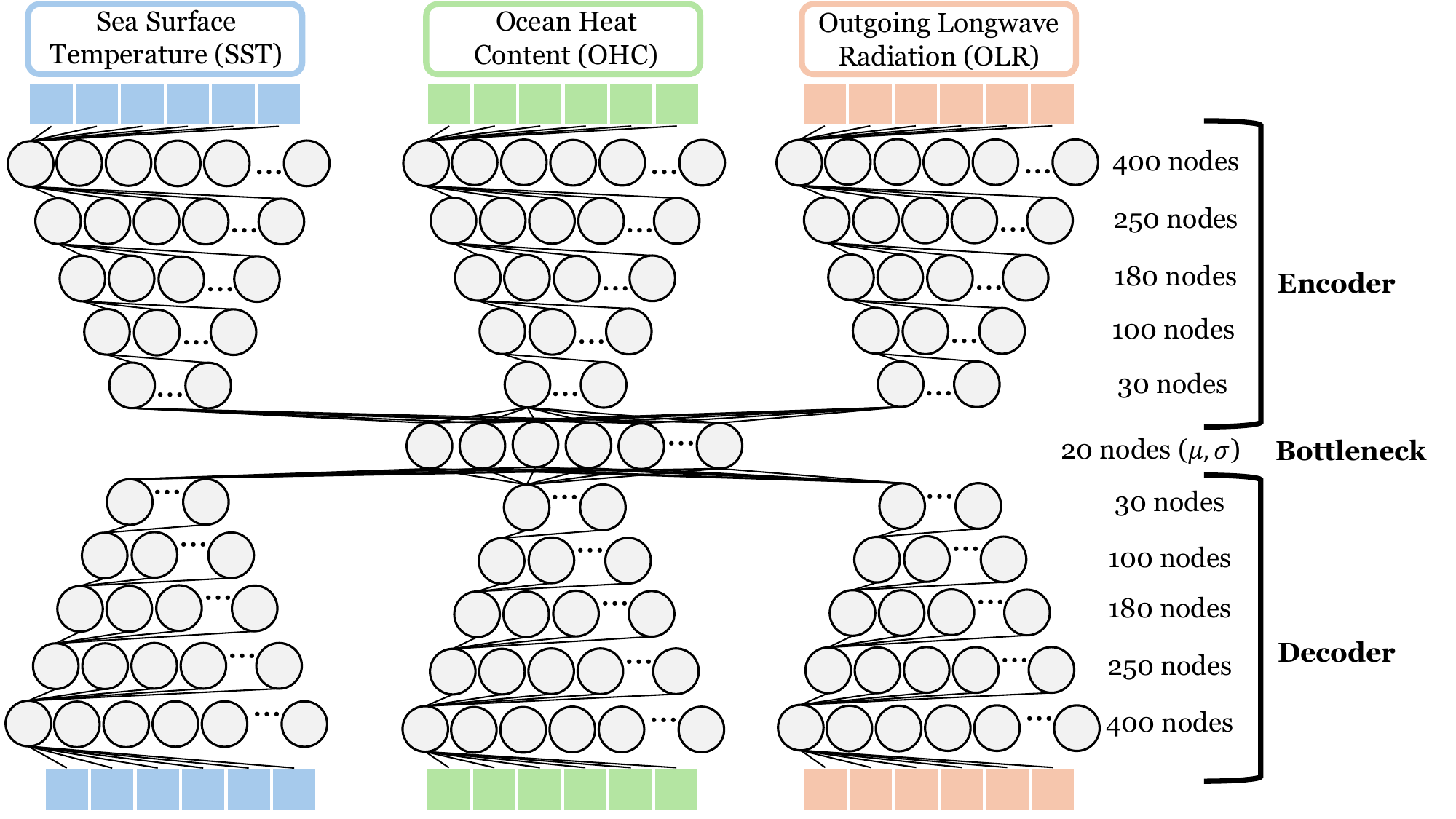}
    \caption{Schematic of the multibranch $\beta$-VAE architecture. For clarity, hidden layer widths are reduced, and only connections to the first node following each input variable are shown. In the implemented model, all neurons are fully connected to the subsequent layer.}
    \label{fig:vae_architecture}
\end{figure*}

Our $\beta$-VAE architecture builds upon the two-branch autoencoder of \cite{passarella2023assessing} by adding generative capabilities and extending the architecture to three variables, each represented by its own encoder-decoder branch (Fig. \ref{fig:vae_architecture}). A manual hyperparameter search was conducted to identify an architecture that achieved high reconstruction performance while maintaining a sufficiently regularized latent space to support latent traversals. The search was first performed for univariate $\beta$-VAEs (SST, OHC, and OLR separately) across optimizers, learning rates, batch sizes, training epochs, validation splits, latent dimensions, model complexity (width and depth), and random seeds. We found that model performance was relatively insensitive to variations in learning rates, but improved with increases in the latent dimensions, hidden layers, and validation sets. Guided by these results, the multivariate $\beta$-VAE search space was constrained to variations in batch sizes, model complexity, training epochs, and validation splits.

The final hyperparameters for the multivariate $\beta$-VAE were a batch size of 256 for stable optimization and early stopping at 150 epochs to limit overfitting (Table \ref{tab:vae}). The hyperbolic tangent was used as an activation function due to its performance during the hyperparameter search and its zero-centered nonlinearity, which is well-suited to both positive and negative inputs. Sensitivity experiments with reduced latent dimensionality ($<10$; see Appendix \ref{sec:dimensionality}) indicated that stronger compression led to latent monopolization (i.e., a small subset of latent dimensions carried most of the information). Therefore, the $\beta$-VAE's latent space was expanded to 20 dimensions (Fig. \ref{fig:vae_architecture}), as compared to the fewer latent dimensions contained in \cite{passarella2023assessing}, to maximize latent space capacity and reconstruction skill. Skill, as defined by our loss function, showed little sensitivity to random weight initialization, suggesting that the final hyperparameter configuration was robust (additional information is located in Appendix \ref{sec:ensemble}). 

The $\beta$-VAE is trained with a loss function that trades off the reconstruction accuracy of the input variables and regularization of the latent space toward a predefined prior distribution. For reconstruction accuracy of each variable, the mean squared error (MSE) is computed between the original input ($X_i$) and its decoded reconstruction ($\hat{X}_i$), where $N$ is the minibatch size:

\begin{equation}
    \text{MSE} = \frac{1}{N} \sum_{i=1}^{N} \left( X_i - \hat{X}_i \right)^2.
\end{equation}

\noindent The total reconstruction loss ($\mathcal{L}_{\text{recon}}$) is the sum of the three variable-specific MSEs: 

\begin{equation}
\mathcal{L}_{\text{recon}} 
= \text{MSE}_{\text{SST}}(X, \hat{X})
+ \text{MSE}_{\text{OHC}}(X, \hat{X})
+ \text{MSE}_{\text{OLR}}(X, \hat{X}).
\end{equation}

\noindent The $\mathcal{L}_{\text{recon}}$ encourages the $\beta$-VAE to reconstruct each variable accurately. Training and validation loss curves are tracked separately for SST, OHC, and OLR to assess whether the $\beta$-VAE was preferentially learning the structure of one variable at the expense of the others (shown in Figure \ref{fig:vae_loss} of the Appendix). 

Latent regularization supports the generative aspect of the $\beta$-VAE by encouraging the distribution of the latent space to remain close to a predefined prior distribution that is independent of any particular input. Latent regularization effectively constrains the capacity of the latent space and promotes generalization to unseen data, thereby enabling subsequent sampling from the latent space. We use the standard multivariate normal as the prior $p(z)=\mathcal{N}(0, I)$, where $I$ is the identity matrix specifying unit variance in each latent dimension and zero covariance between distinct latent dimensions. The encoder outputs the parameters of a distribution ($\mu, \sigma$) that represents possible latent representations $z_i$ for $X_i$. This distribution, the encoder's approximate posterior $q_\phi(z_i|X_i)$, is modeled as a 20-dimensional diagonal Gaussian, 

\begin{equation}
    q_\phi(z_i|X_i)=\mathcal{N}(\mu_\phi(X_i), \text{diag}(\sigma_\phi^2(X_i))),
\end{equation}
\noindent where $\phi$ denotes the encoder parameters (weights and biases), and $\mu_\phi(X_i)$ and $\sigma_\phi(X_i)$ are the encoder outputs (means and standard deviations) for each sample per latent dimension. Thus, we obtain sample specific $z_i$ values from the encoder $z_i\sim q_\phi(z_i|X_i)$ or $z_i=\mu_\phi(X_i)$ and decode them to obtain $\hat{X}_i$. During data generation, we can draw $z\sim p(z)$ and pass $z$ through the decoder to produce a synthetic sample $\hat{X}$, where $z$ represents the latent vector sampled from the prior distribution $p(z)$. 

The Kullback-Leibler divergence term \citep[KL divergence;][]{odaibo2019tutorial} measures the discrepancy between the encoder’s approximate posterior $q_\phi (z_i|X_i)$ and the prior $p(z)$. KL divergence can be computed in closed form as:

\begin{equation}
\mathcal{L}_{\text{KL}} = -\tfrac{1}{2} \sum_{j=1}^{20} \left( 1 + \log \sigma_j^2 - \mu_j^2 - \sigma_j^2 \right),
\end{equation}

\noindent where $j$ corresponds to the latent dimension. The total loss minimized during training is defined as the weighted sum of the reconstruction loss and the KL divergence, following the $\beta$-VAE formulation in \cite{higgins2017betavae}:

\begin{equation}
\mathcal{L}_{\text{total}} = \mathcal{L}_{\text{recon}} + \beta \, \mathcal{L}_{\text{KL}},
\end{equation}

\noindent where the $\mathcal{L}_{\text{KL}}$ term is scaled by a small weighting factor of choice; we use $\beta = 0.0005$. This weighting scheme prioritizes reconstruction fidelity while substantially reducing the relative influence of the KL-divergence term that encourages the approximate posterior to remain close to the prior. Several larger $\beta$ values, between $0.01$ and $1$, were evaluated and found to concentrate information in a smaller set of nodes. Therefore, the small weighting factor of $0.0005$ was selected to preserve reconstruction fidelity across latent dimensions while maintaining regularization of the latent space.

A challenge when training a $\beta$-VAE with backpropagation is that sampling $z_i$ from a distribution $q_\phi(z_i \mid X_i)$ is non-differentiable with respect to the encoder parameters $\phi$. To enable gradient-based optimization, the reparameterization trick \citep{kingma2013auto} is employed, which expresses the latent variable as: 

\begin{equation}
    z_i = \mu_\phi(X_i) + \exp(0.5 \cdot \log\sigma^2_{\phi}(X_i)) \odot \epsilon_i, 
\quad \epsilon_i \sim \mathcal{N}(0, I),
\end{equation}

\noindent where $\mu_\phi \in \mathbb{R}^{d}$ is the latent mean, $d = 20$ number of latent dimensions, and
$\log(\sigma_\phi^2) \in \mathbb{R}^{d}$ is the log-variance output by the encoder. The standard deviation is recovered as $\sigma_\phi = \exp(0.5\log \sigma_\phi^2)$, which is a numerically stable way to enforce positive variance. The noise vector $\epsilon_i \in \mathbb{R}^{d}$ is sampled from a standard normal distribution. The resulting latent vector $z_i\in\mathbb{R}^{d}$ is then passed through the decoder to reconstruct the input $\hat{X}_i$.

As training progressed, the KL divergence term increased as a function of epoch, mirroring the decline in reconstruction losses (Fig. \ref{fig:vae_loss}). The KL divergence term plateaued after 80 epochs, indicating that the multivariate $\beta$-VAE had largely converged under the chosen objective.

\subsection{Latent Variance Distribution}\label{lvardist}

To assess whether the $\beta$-VAE effectively uses its latent space, we quantify how much the encoder posterior means vary across the dataset, thereby identifying potentially inactive latent dimensions. For each input sample $X_i$, the encoder defines an approximate posterior $q_\phi(z_i|X_i)$ and outputs its mean vector $\mu_\phi(X_i) \in \mathbb{R}^d$. We use this posterior mean as a deterministic latent representation and define

\begin{equation}
    z_i \equiv \mu_\phi(X_i) \in \mathbb{R}^{d},\,\,\, i=1,\dots,N,
\end{equation}

\noindent where $N=6000$ is the total number of training and test samples. Let $z_{ij}$ denote the $j$-th component of $z_i$ (latent dimension $j$), with $j=1, \dots, d$ and $d = 20$. Utilization is assessed by measuring the variability of $z_{i,j}$ across samples $i$ for each fixed $j$. 

We compute the per-dimension variance by first calculating the dataset mean latent vector $\bar{z}$,

\begin{equation}
\bar{z} = \frac{1}{N} \sum_{i=1}^{N} z_i,
\end{equation}

\noindent and use the resultant $\bar{z}$ to center each latent vector $z_i$ as follows

\begin{equation}
\tilde{z}_i = z_i - \bar{z}, \quad i = 1, \dots, N.
\end{equation}

\noindent The empirical variance of latent dimension $j$ across the dataset $N=6000$ is then

\begin{equation}
\mathrm{Var}(z_j) = \frac{1}{N} \sum_{i=1}^{N} \left( \tilde{z}_{ij} \right)^2.
\end{equation}

\noindent To compare dimensions on a common scale, we compute the normalized variance share:

\begin{equation}
\text{VarShare}_j = \frac{\mathrm{Var}(z_j)}{\sum_{k=1}^{d} \mathrm{Var}(z_k)}.
\end{equation}

\begin{figure}[t]
    \centering
    \includegraphics[width=34pc]{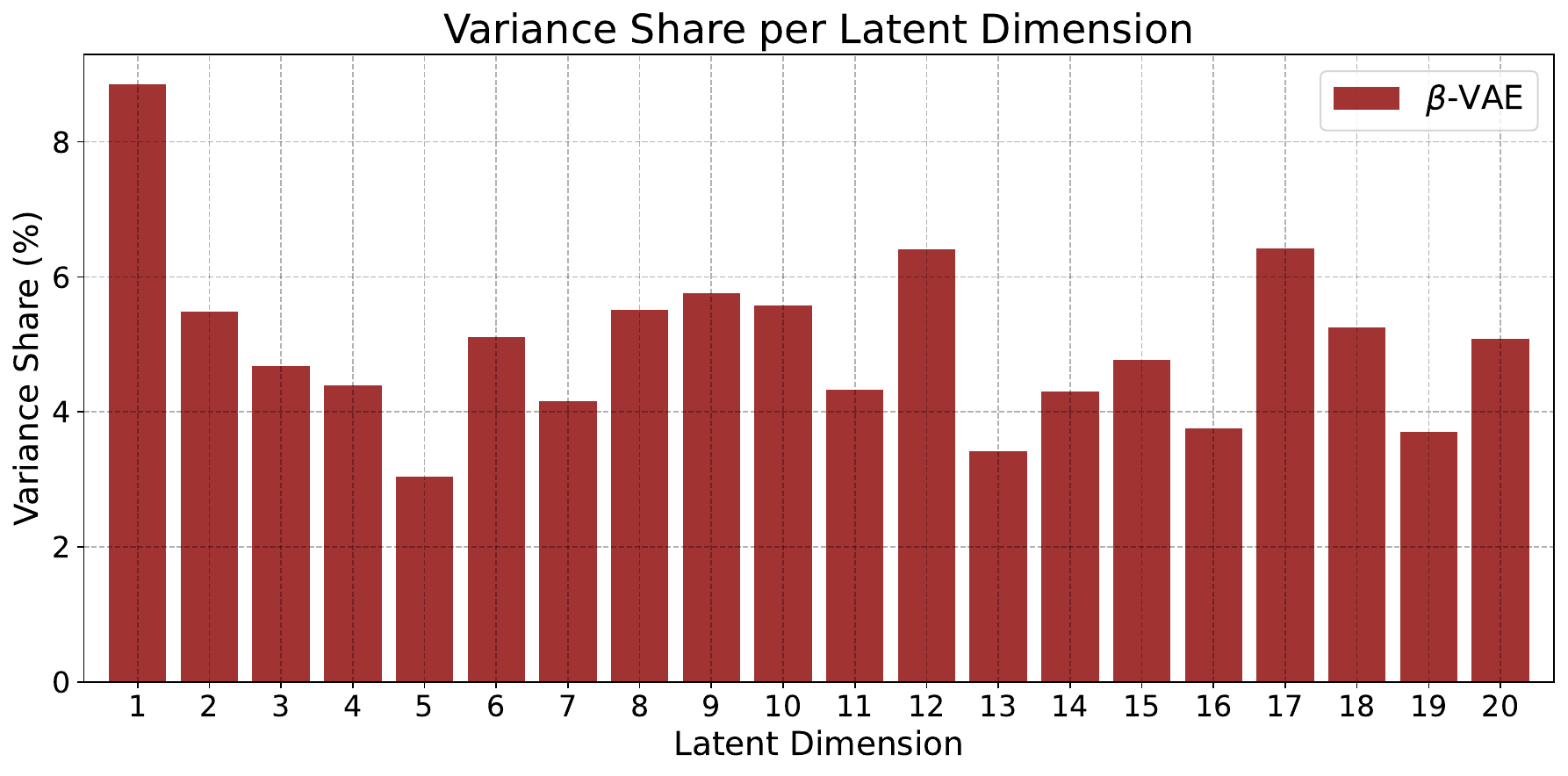}
    \caption{The variance share from each latent dimension of the multi-branch $\beta$-VAE derived from the centered latent vectors, summing to 100\%.}
    \label{fig:variance_explained}
\end{figure}

\noindent The vector $\text{VarShare} \in \mathbb{R}^{d}$ ($d=20$) summarizes the proportion of the total marginal latent variance attributable to each latent dimension. Latent dimensions with near-zero variance $\mathrm{Var}(z_j)$ exhibit little change across inputs and suggest redundancy or under-utilization of that latent dimension. This phenomenon is often discussed in the context of posterior collapse, in which the encoder's approximate posterior $q_\phi(z|X)$ becomes close to the prior $p(z)$ and therefore carries little information about $X$ \citep{wang2023posterior}. Figure \ref{fig:variance_explained} shows that the variance shares (expressed as percentages) are relatively uniform across all 20 latent dimensions, with each dimension accounting for between 3\% and 9\% of the total variance. This result suggests that the encoder utilizes all latent dimensions of the $\beta$-VAE and that posterior collapse has not occurred.

\section{Results}
\subsection{Multi-branch $\beta$-VAE Reconstruction Skill}

For the three variables at each grid cell, we assessed the $\beta$-VAE's reconstruction skill along the dataset's time dimension $i$ using mean absolute error (MAE), root mean squared error (RMSE), and coefficient of determination ($R^2$); see Appendix for more details. RMSE and MAE summarize the magnitude of reconstruction errors: RMSE emphasizes large errors and is sensitive to outliers, whereas MAE measures absolute deviation. $R^2$ quantifies the fraction of spatial variance at each time step explained by the reconstruction. To assess the extent of skill degradation between training and test sets, we compute skill ratios for MAE, RMSE, and $R^2$. The skill ratio (SR) is computed as: 

\begin{equation}
    \text{SR}=\frac{\text{Test}}{\text{Train}},
\end{equation}

\noindent where `Test' represents the corresponding test set metric result and `Train' represents the corresponding training set metric result. For MAE and RMSE, $\text{SR}>1$ represents skill degradation in the test set as compared to the training set performance (error ratio). For $R^2$, $\text{SR}<1$ represents reduced variance explained in the test set as compared to the training set (retention ratio).

To estimate sampling variability in the difference between train and test set performance, we applied a bootstrap resampling along the time dimension. Specifically, we generated $B=1000$ bootstrap replicates by independently drawing test ($N_{\text{test}}=1200$) and training data ($N_{\text{train}}=4800$) time indices with replacement, constructing the corresponding resampled time series, and computing error metrics as described above. For each bootstrap replicate, differences between the test and training metrics were used to obtain the resulting distribution of differences. Confidence intervals that include zero indicate no statistically significant difference between the train and test set performance. For MAE and RMSE, intervals entirely above zero indicate poorer performance on the test set, whereas for $R^2$, intervals entirely below zero indicate poorer test-set performance. 
 
\begin{figure*}[t]
    \centering
    \includegraphics[width=1\linewidth]{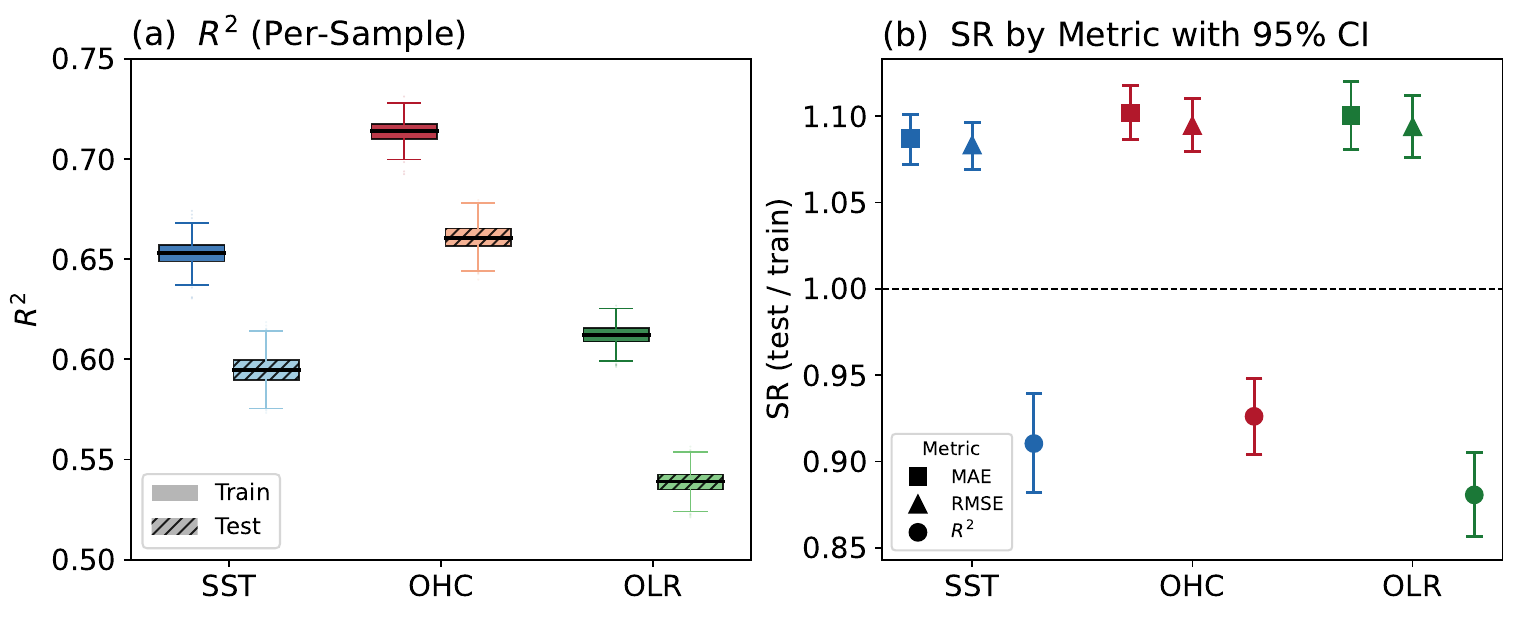}
    \caption{(a) Bootstrap distributions of $R^2$ ($B=1000$ replicates, $N=1200$ samples) for training (solid) and testing (hatched) data for SST, OHC, and OLR. (b) Skill ratios (SR = test/train) with $95\%$ confidence intervals for MAE (squares), RMSE (triangles), and $R^2$ (circles). The horizontal dashed line at $\text{SR}=1$ indicates equal performance between training and testing sets. $\text{SR}>1$ for MAE and RMSE, and $\text{SR}<1$ for $R^2$, indicates degraded test performance relative to training.}
    \label{fig:r2_sr}
\end{figure*}

Figure \ref{fig:r2_sr} summarizes $\beta$-VAE reconstruction skill across training and testing datasets. $R^2$ exceeds 0.5 in both training and testing, and is lowest for OLR, likely reflecting the $\beta$-VAE's ability to capture variance in slower-evolving oceanic fields compared to higher-frequency atmospheric variability. MAE and RMSE SRs range from $1.08-1.10$ and $R^2$ SRs from $0.88-0.93$. This indicates a small but consistent generalization gap across all input variables. Detailed error statistics are outlined in Table \ref{tab:metrics} of the Appendix.

\begin{figure*}[t]
    \centering
    \includegraphics[width=21pc]{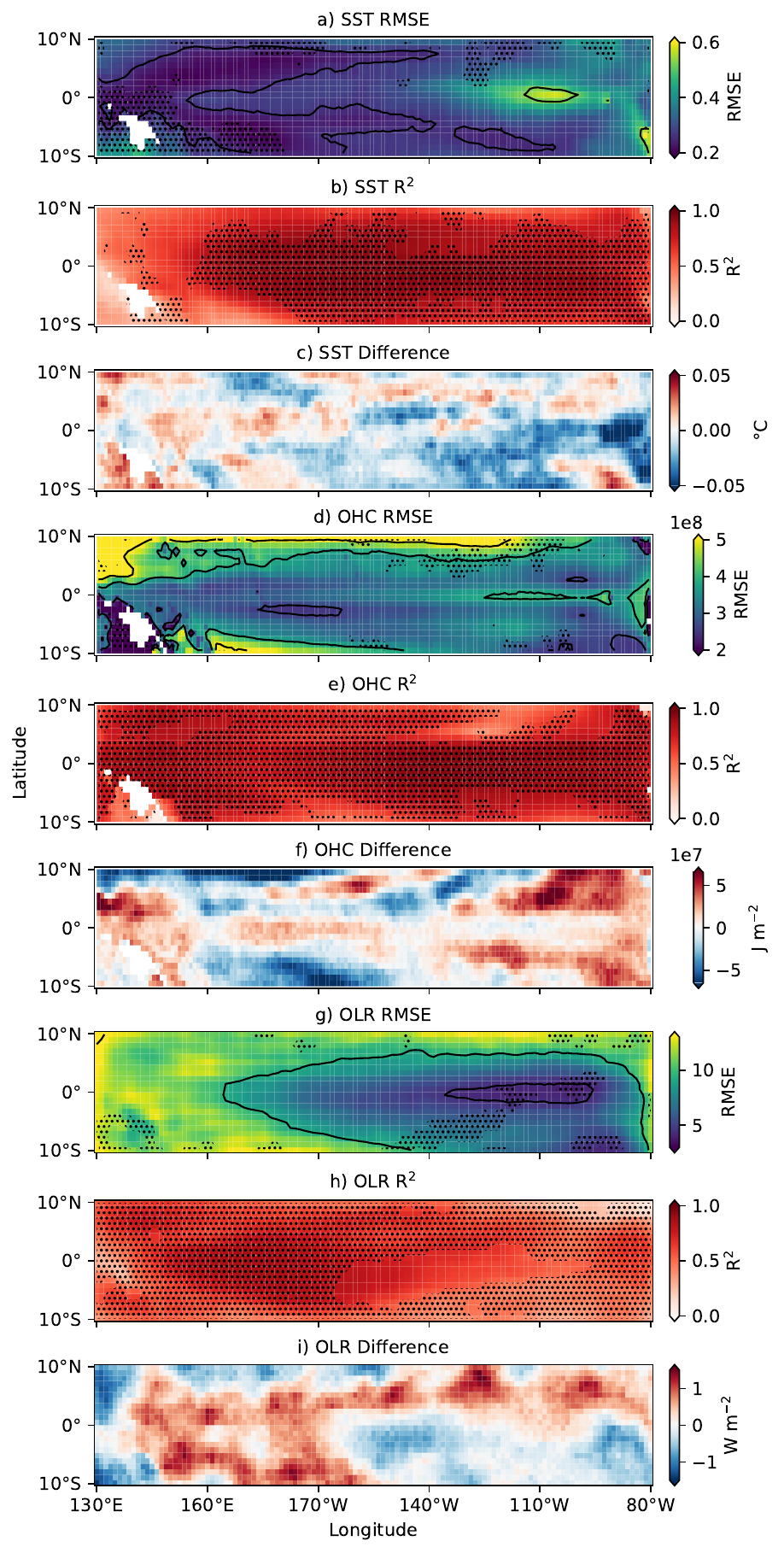}
    \caption{RMSE, MAE (black contours), $R^2$, and differences (original minus reconstructed) for SST ($^{\circ}C$, a-c), OHC ($\times10^{8}J\,m^{-2}$, d-f), and OLR ($W\,m^{-2}$, g-i). Three evenly spaced MAE contours span $0.2–0.6^{\circ}C$, $2-4\times10^{8}\ J\,m^{-2}$, and $3-11\ W\,m^{-2}$, respectively. Stippling indicates test-set performance within the one-sided 95\% training-bootstrap threshold for RMSE and MAE, and 5\% training-bootstrap threshold for $R^2$ ($B=1000$).}
    \label{fig:spatial_metrics}
\end{figure*}

We then assess spatial patterns of reconstruction skill by computing MAE, RMSE, and $R^2$ per valid grid cell over time steps. An upper confidence bound was estimated for each grid cell by resampling (with replacement; $B=1000$) paired input fields and reconstructions from the training set (using $N_\text{train}=1200$), computing performance metrics, and deriving the upper confidence bound from the bootstrap distribution of these metrics. The upper bound was used for MAE and RMSE ($\geq$95\textsuperscript{th} percentile), such that test-set errors exceeding this threshold were flagged as statistically significantly `worse' than the training set. For $R^2$, a complementary analysis was conducted using the lower confidence bound ($\leq$5\textsuperscript{th} percentile of $R^2$) from a 1000-member bootstrap ($B=1000$). If the test set $R^2$ was below the lower confidence bound, the value was flagged as statistically significantly `worse' compared to the training set $R^2$ at that grid cell. 24- and 36-month block bootstrap windows were also used to assess potential sensitivity to temporal autocorrelation, yielding results consistent with our implemented approach (not shown). Stippling in Figure \ref{fig:spatial_metrics} denotes regions where the test-set performance skill is within the training set bootstrap distribution at that grid cell (not thus, not statistically significantly `worse' than the training set). Lastly, to identify where the $\beta$-VAE reconstructions deviate from the input fields, Figure \ref{fig:spatial_metrics}(c,f,i) highlights the spatial structure of local reconstruction errors for the test set. Red (shaded) colors indicate regions where the $\beta$-VAE reconstruction is higher on average, while blue (shaded) colors correspond to regions where it is lower.

For SST, RMSE is lowest in the western tropical Pacific and increases towards the central and eastern basin, with the largest errors occurring from approximately $120^{\circ}$W to $100^{\circ}$W at the equator and in the southeastern portion of the domain at $10^{\circ}$S (Figure \ref{fig:spatial_metrics}a). MAE contours largely follow the shaded RMSE patterns. Regions of stippling are found south of the equator in the western tropical Pacific and north of the equator in the eastern tropical Pacific near $130^{\circ}$W, which indicates that test performance is consistent with training performance based on the $95$th percentile confidence level (Figure \ref{fig:spatial_metrics}a). In contrast, the highest $R^2$ is in the central and eastern tropical Pacific, while lower $R^2$ is in the western basin and off-equatorial regions (Figure \ref{fig:spatial_metrics}b). Stippling is pronounced across much of the central and eastern basin east of approximately $160^{\circ}$E (Figure \ref{fig:spatial_metrics}b). SST reconstructions exhibit a weak negative bias in the equatorial cold tongue (Figure \ref{fig:spatial_metrics}c). The error and variance patterns highlight an east-west contradiction in SST reconstruction performance: large RMSE and MAE coincide with higher $R^2$, and small RMSE and MAE coincide with lower $R^2$. The apparent contradiction arises from the magnitude of SST anomalies in the eastern and western Pacific. Regions with larger ENSO-driven SST anomalies (i.e., the central and eastern Pacific) can exhibit larger absolute reconstruction differences while still explaining a greater fraction of their temporal variance. This $\beta$-VAE performance aligns with the known structure of tropical Pacific SST variability, in which the central and eastern basin is dominated by large-amplitude and spatially coherent ENSO anomalies, while SST anomalies in the western Pacific warm pool are of weaker magnitude compared to the central and eastern basin \citep{trenberth1997definition, neelin1998enso, deser2010sea, capotondi2015understanding}.

RMSE and MAE are larger for OHC over the off-equatorial bands of the tropical Pacific (Figure \ref{fig:spatial_metrics}d). Generally, errors are of smaller magnitude along the equator, with the lowest errors encompassing the windward side of Papua New Guinea from $130^{\circ}$E to $140^{\circ}$E. This error structure is consistent with dynamical processes governing OHC across the equatorial thermocline, which exhibits a basin-scale coherence associated with ENSO variability \citep{zebiak1987model, meinen2000observations}. There are a few regions where test performance is consistent with training performance, including on the windward side of Papua New Guinea and off-equatorial regions between $170^{\circ}$W and $100^{\circ}$W, indicated by regions of stippling (Figure \ref{fig:spatial_metrics}d). Lower $R^2$ near $5^{\circ}$N and $125^{\circ}$W (Figure \ref{fig:spatial_metrics}e) is potentially influenced by a range of dynamically active off-equatorial processes, including subsurface variability associated with the North Equatorial Undercurrent and wave-driven thermocline variability associated with off-equatorial Rossby waves \citep{chelton1996global, johnson1999interior, chen2016interannual}. These subsurface processes can modulate OHC anomalies while not being visible on the overlying sea surface \citep{deser2010sea}. The $\beta$-VAE appears to explain more OHC variance (higher $R^2$) in regions where subsurface OHC anomalies have a stronger surface expression, such as the equator (stippling), and lower where subsurface variability is weakly expressed at the surface. Compared to the original OHC inputs, OHC reconstructions are larger in magnitude in the eastern Pacific (Figure \ref{fig:spatial_metrics}f).

For OLR (Figure \ref{fig:spatial_metrics}g,h), errors are largest in the western tropical Pacific, coinciding with regions of enhanced convection associated with the warm pool \citep{waliser1993comparison}. Stippling is located in the western Pacific near Papua New Guinea, and again in the central and eastern basin from approximately $150^{\circ}$W to $80^{\circ}$W, encompassing both equatorial and off-equatorial regions (Figure \ref{fig:spatial_metrics}g). In contrast to the slowly evolving oceanic components, $R^2$ values for OLR are comparatively lower, particularly in the eastern equatorial Pacific, where eastward-propagating, high-frequency atmospheric variability associated with tropical convection occurs and strongly modulates OLR \citep{zhang2005madden, kiladis2009convectively}. $R^2$ stippling is pronounced across much of the domain, except for the equatorial region from $160^{\circ}$W to $90^{\circ}$W (Figure \ref{fig:spatial_metrics}h). OLR reconstructions exhibit drier conditions in the western and central Pacific (Figure \ref{fig:spatial_metrics}i).

The spatial patterns of the performance metrics are broadly consistent with known tropical Pacific dynamics. Notably, regions with relatively low RMSE can still exhibit low $R^2$ where variance is small (e.g., warm pool SSTs), whereas regions with larger variability may show larger $R^2$ despite larger absolute errors (e.g., eastern Pacific SSTs). In contrast, other results indicated the concurrence of high errors and low explained variance (e.g., OHC at $5^{\circ}$N, $125^{\circ}$W). Regions exhibiting low errors and high $R^2$ indicate grid cells in which anomalies are well reconstructed and well captured in both amplitude and phase by the $\beta$-VAE (e.g., equatorial OHC).

\subsection{Latent Space Variance Diagnostics}
\label{variance_diagnostics}

\begin{figure*}[t]
    \centering
    \includegraphics[width=32pc]{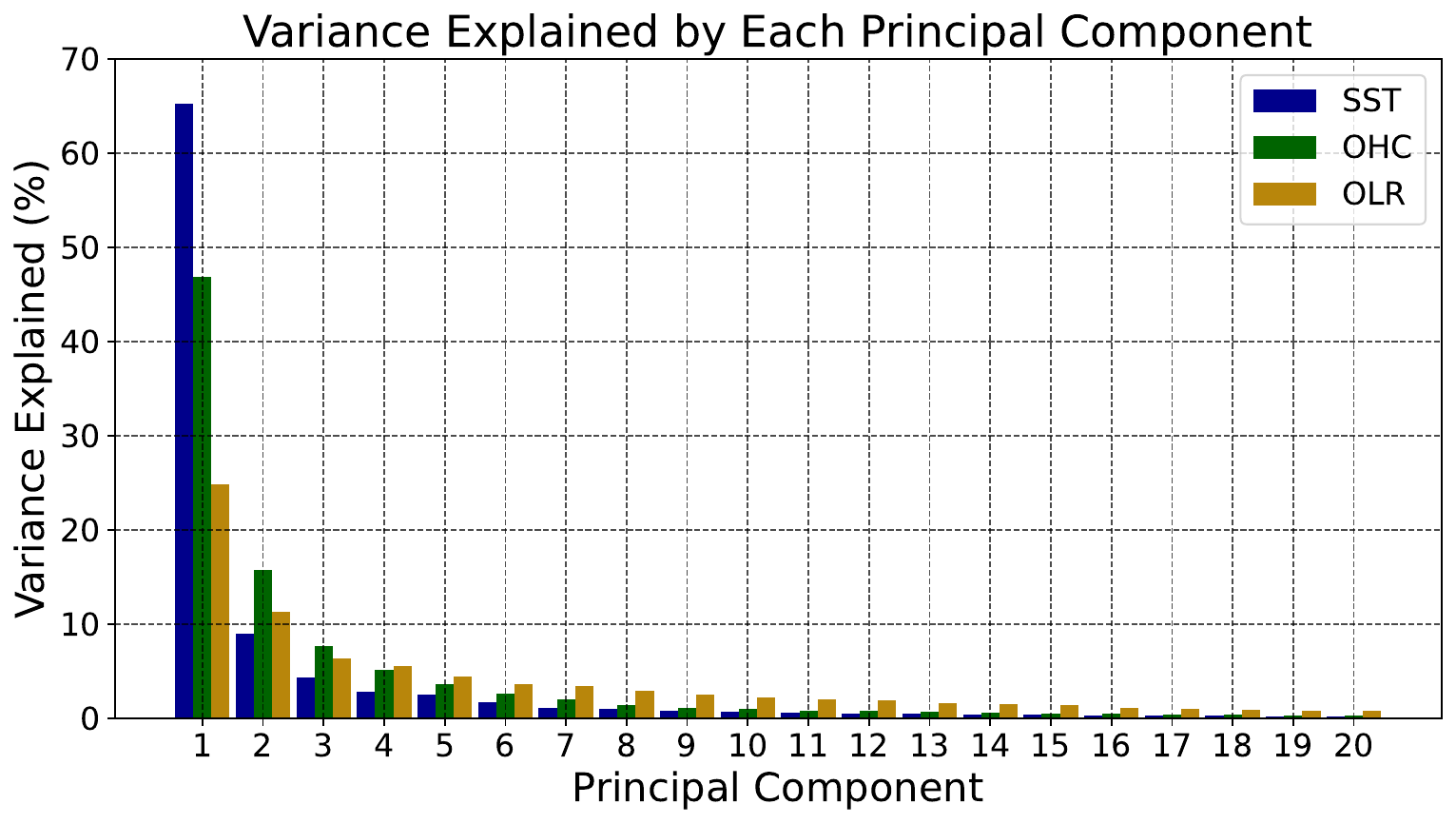}
    \caption{Percentage of variance explained by the first 20 principal components (PCs) derived separately for SST, OHC, and OLR, as legend indicated.}
    \label{fig:pca}
\end{figure*}

ENSO variability is commonly characterized using leading principal components of SSTs, and we therefore apply Principal Component Analysis (PCA) to each original input field (SST, OHC, and OLR) independently to establish a linear, orthogonal dimensionality-reduction reference. For each variable, we retain the first 20 principal components (PCs) and compute the explained variance of each PC, which quantifies how the total variance of that field is distributed across 20 linear modes. Explained variance differs substantially across input variables (Figure \ref{fig:pca}). For SST, the three leading PCs account for approximately $80\%$ of the total variance, suggesting that SST variability is dominated by a small number of large-scale, coherent modes. OHC exhibits a broader distribution, with the first five PCs required to capture a comparable fraction of variance. In contrast, OLR exhibits a more diffuse variance structure, where the eight leading PCs explain approximately $60\%$ of the variance, consistent with higher-frequency, spatially heterogeneous convective variability.

Direct PCA-style explained variance is not well-defined for the $\beta$-VAE latent space. The encoder--decoder mapping is nonlinear and probabilistic, and the latent dimensions $j=1,\dots,d$ are neither orthogonal nor ordered by variance. As a result, variability in the reconstructed fields cannot be cleanly partitioned into additive, non-overlapping `explained variance' contributions per latent dimension, and a cumulative variance curve analogous to PCA is not meaningful because information may be shared across dimensions and expressed through interactions in the decoder. Instead, to compare the $\beta$-VAE latent space to the PCA reference, we use targeted reconstruction and sensitivity experiments that quantify how decoded variability changes when individual latent dimensions are perturbed or isolated. Encoder-side latent utilization was previously assessed using the marginal variance of the latent posterior means (Section 2\ref{lvardist}; Fig. \ref{fig:variance_explained}), which addressed whether latent dimensions are active, but this is not equivalent to variance attribution in the reconstructed fields performed here. We implement two complementary reconstruction experiments to better understand how information is distributed across the nonlinear $\beta$-VAE latent space.

\begin{figure*}[t]
    \centering
    \includegraphics[width=1\linewidth]{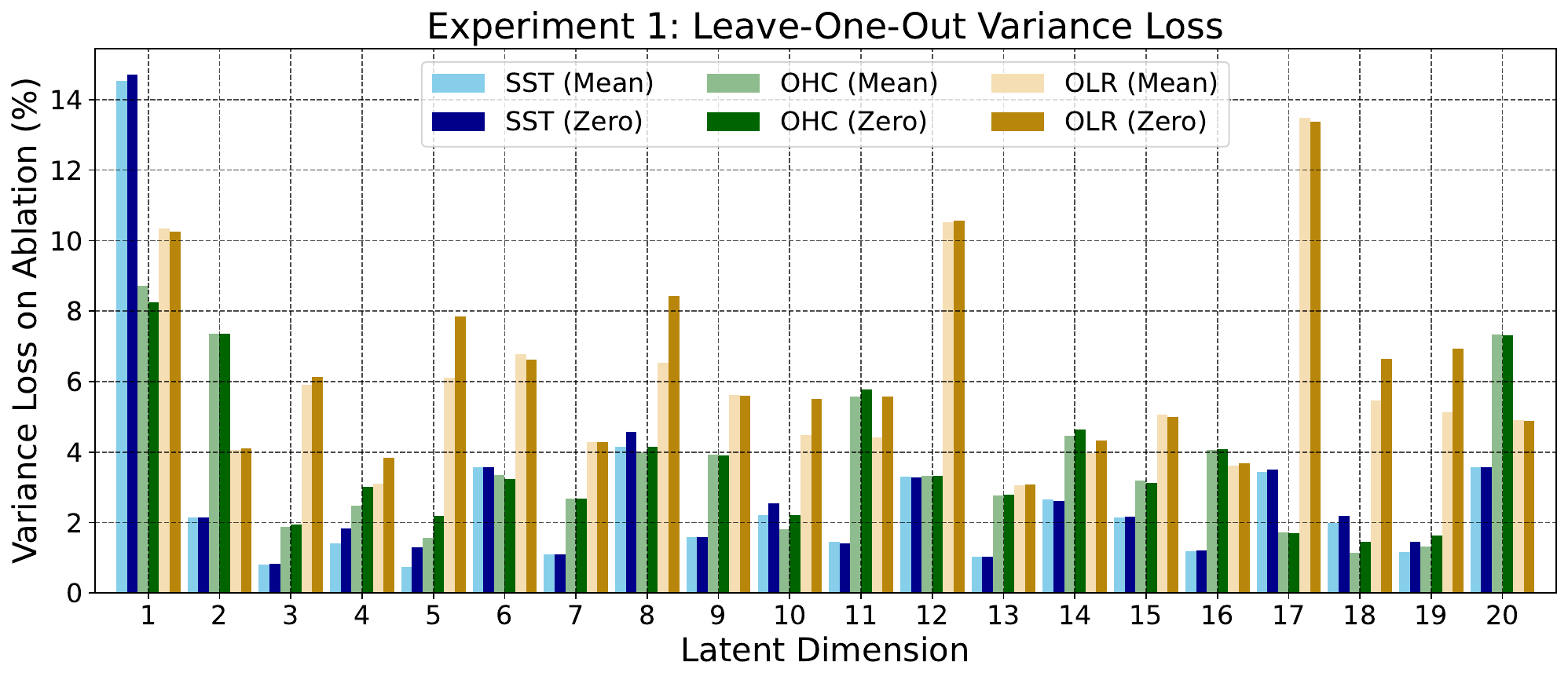}
    \caption{Experiment 1: Leave-one-out reconstruction sensitivity. Bars show the normalized spatial variance of the difference between the full reconstruction and the reconstruction after ablating latent dimension $j$ (by mean- or zero-replacement), expressed as a percentage of the total spatial variance of the full $\beta$-VAE reconstruction for each input variable (legend indicated).}
    \label{fig:exp1_vae}
\end{figure*}

Experiment 1 assesses reconstruction sensitivity to the removal of individual latent dimensions using a leave-one-out framework. For each sample $X_i$, we encode the latent posterior mean $\mu_\phi(X_i) \in \mathbb{R}^d$, with components $\mu_{i,j}, j=1,\dots,d$. For each latent dimension $j$, we form an ablated latent vector by replacing $\mu_{i,j}$ with either its temporal mean or zero, while retaining all other components. The impact of removing latent dimension $j$ is quantified as the spatial variance of the difference between the full reconstruction and the leave-one-out reconstruction, normalized by the total spatial variance of the full $\beta$-VAE reconstruction for each input variable. This procedure is repeated for all $j=1,\dots,d$ (Figure \ref{fig:exp1_vae}).

To identify latent dimensions that contribute disproportionately to the reconstruction, we calculate the mean and standard deviation across latent dimensions from the average of the mean- and zero-replacement ablation-induced variance loss, separately for each input variable. An LD is flagged when the variance loss exceeds the respective mean plus one standard deviation. Thresholds of $5.70\%$, $5.72\%$, and $8.42\%$ are identified for SST, OHC, and OLR, respectively. For SST, reconstruction sensitivity is concentrated in the first latent dimension, LD1 ($j=1$), with both mean and zero replacements yielding approximately $14\%$ loss of normalized variance. All remaining latent dimensions contribute substantially less, with individual sensitivities between $1-4\%$. This result suggests that SST-related decoded variability is largely conditioned on a small subset of latent dimensions, whereas the remaining dimensions exhibit comparatively weak sensitivity when removed individually. OHC exhibits a broader distribution of reconstruction sensitivity across latent dimensions, where LD1, LD2, and LD20 exceed the threshold. All remaining dimensions contribute approximately $2-5\%$. This result suggests that OHC-related decoded variability depends on a larger subset of latent dimensions than SST. For OLR, LD1, LD12, and LD17 exceed the threshold, with LD17 contributing the largest effect at approximately $14\%$. This pattern is consistent with OLR's spatially heterogeneous variability and suggests that OLR information is distributed across multiple interacting latent coordinates.

\begin{figure*}[t]
    \centering
    \includegraphics[width=1\linewidth]{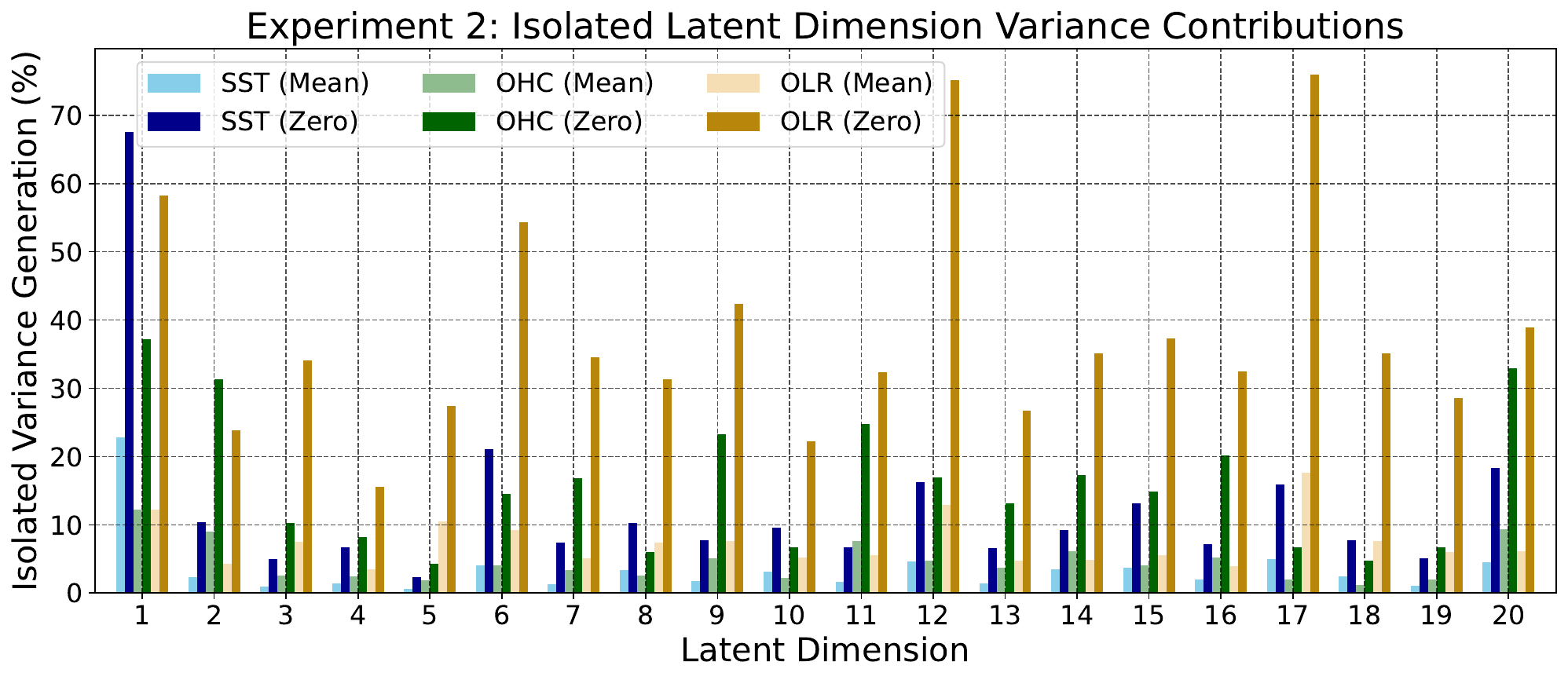}
    \caption{Same as Figure \ref{fig:exp1_vae}, but for Experiment 2: Isolated latent-dimension variance generation. For each latent dimension $j$, reconstructions are decoded while holding all other latent dimensions fixed to a baseline (mean or zero).}
    \label{fig:exp2_vae}
\end{figure*}

Experiment 2 quantifies the variability that individual latent dimensions can generate in isolation. For each sample $X_i$, we allow a single latent dimension $j$ to take its encoded value while all remaining dimensions are held fixed to a baseline state. As in Experiment 1, two baselines are considered: the temporal mean latent vector and the zero vector. For each latent dimension $j$, we decode the resulting latent vector and compute the spatial variance of the reconstruction, normalized by the total spatial variance of the full $\beta$-VAE reconstruction for the corresponding input variable and expressed as a percentage (Figure \ref{fig:exp2_vae}).

For SST, isolated variance generation is dominated by LD1, consistent with Experiment 1; LD1 alone accounts for nearly $70\%$ of the reconstructed spatial variance under the zero baseline, whereas the mean baseline yields approximately $20\%$. All remaining latent dimensions generate substantially less SST variance in isolation, with LD6 and LD20 producing moderate isolated variance under the zero baseline at approximately $20\%$, and remaining dimensions generating approximately $5\%-15\%$ across both baselines. These results indicate that a small subset of latent dimensions can generate a large fraction of the SST variability when activated in isolation. OHC exhibits a more distributed pattern of isolated variance generation across the latent space. While LD1 remains influential (nearly $40\%$) under the zero baseline, LD2, LD9, LD11, and LD20 each generate more than $20\%$ for OHC. Under the mean baseline, isolated variance is substantially smaller, with LD1 and LD20 producing only about $10\%$. In contrast to SST and OHC, OLR displays the widest range of influential latent dimensions. LD1, LD6, LD12, and LD17 generate appreciable OLR variance in isolation under the zero baseline ($>50\%$).

Across Experiments 1 and 2, several latent dimensions are repeatedly identified as influential, most notably LD1 for all variables, and LD12 and LD17 for OLR, underscoring that latent dimensions can be important both in terms of reconstruction sensitivity (Experiment 1) and in generating isolated variance (Experiment 2). Finally, sensitivities are broadly similar between mean and zero replacement in Experiment 1, but differences in Experiment 2 suggest that the zero baseline can place the decoder farther from the typical latent manifold associated with the data, thereby amplifying the variability of individual latent dimensions.

\subsection{Interpreting Latent Dimensions using Known ENSO Characteristics}

\begin{figure*}[t]
    \centering
    \includegraphics[width=1\linewidth]{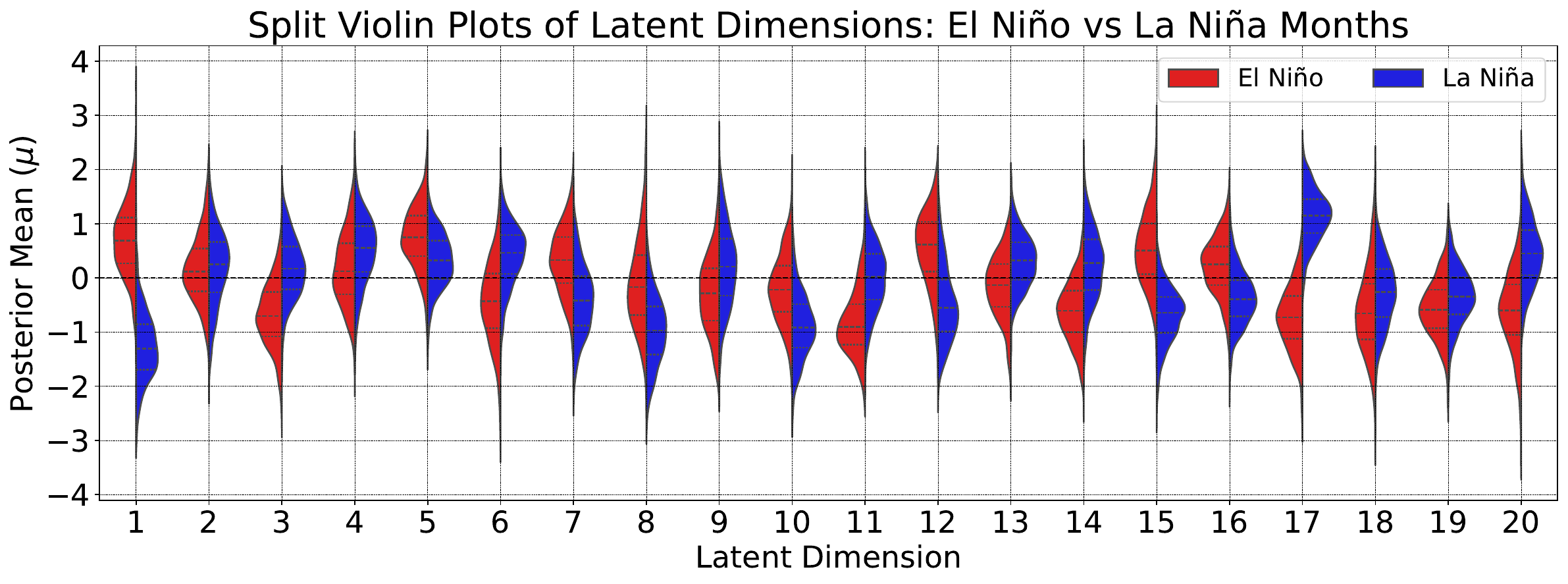}
    \caption{Split violins of the latent posterior means ($\mu_\phi$) for ONI-defined El Niño (red, left) and La Niña months (blue, right). Width indicates distribution density, and dashed lines mark the 25\textsuperscript{th}, median (50\textsuperscript{th}), and 75\textsuperscript{th} percentiles within each ENSO phase.}
    \label{fig:violin}
\end{figure*}

To assess whether the latent space separates ONI-defined El Niño and La Niña conditions, we analyze the latent posterior means ($\mu_\phi$) for each encoded input month. Using ONI computed from the original monthly SSTs from the E3SMv2 piControl ($N=6000$), we identify 831 El Niño and 819 La Niña months. Neutral months ($N=4350$) are omitted to better visualize the active phases of ENSO. LD1, LD17, and LD20 show limited overlap based on El Niño and La Niña, indicating that the $\beta$-VAE encodes variability across SST, OHC, and OLR fields associated with ONI-based ENSO (Figure \ref{fig:violin}). On the other hand, LD2, LD4, and LD13 show substantial overlap between ONI-defined active ENSO phases, suggesting that these latent dimensions may capture information unrelated to ENSO or encode ENSO-related information that is missing from ONI.

To further assess how ONI-defined El Niño and La Niña events map onto latent dimensions, we derive the longitude of peak equatorial SST anomaly for each ENSO event (not month) and relate it to latent posterior means ($\mu_\phi$). `Events' are defined as at least 5 consecutive months of the same ENSO phase, yielding 92 El Niño and 82 La Niña events with mean durations of 7.7 and 8.3 months, respectively. We further subset EP and CP events using a longitude threshold of $140^{\circ}$W, producing 44 CP- and 48 EP-type El Niño events, and 30 CP- and 52 EP-type La Niña events. To obtain the longitude of maximum SST, monthly SST anomalies are averaged over $3^{\circ}$S$-3^{\circ}$N and then grouped by event. This narrow equatorial band is used to reduce potential off-equatorial influences. The longitude of peak SST anomaly is a standard diagnostic of ENSO diversity: event anomalies peaking near the dateline are associated with CP-type, whereas event anomalies peaking farther east, typically the Niño-3 region ($150^{\circ}$W$-90^{\circ}$W) and in stronger cases extending toward Niño-1+2 ($90^{\circ}$W$-80^{\circ}$W), are associated with EP-type \citep{kao2009contrasting, yang2018nino}.

\begin{figure*}[t]
    \centering
    \includegraphics[width=1\linewidth]{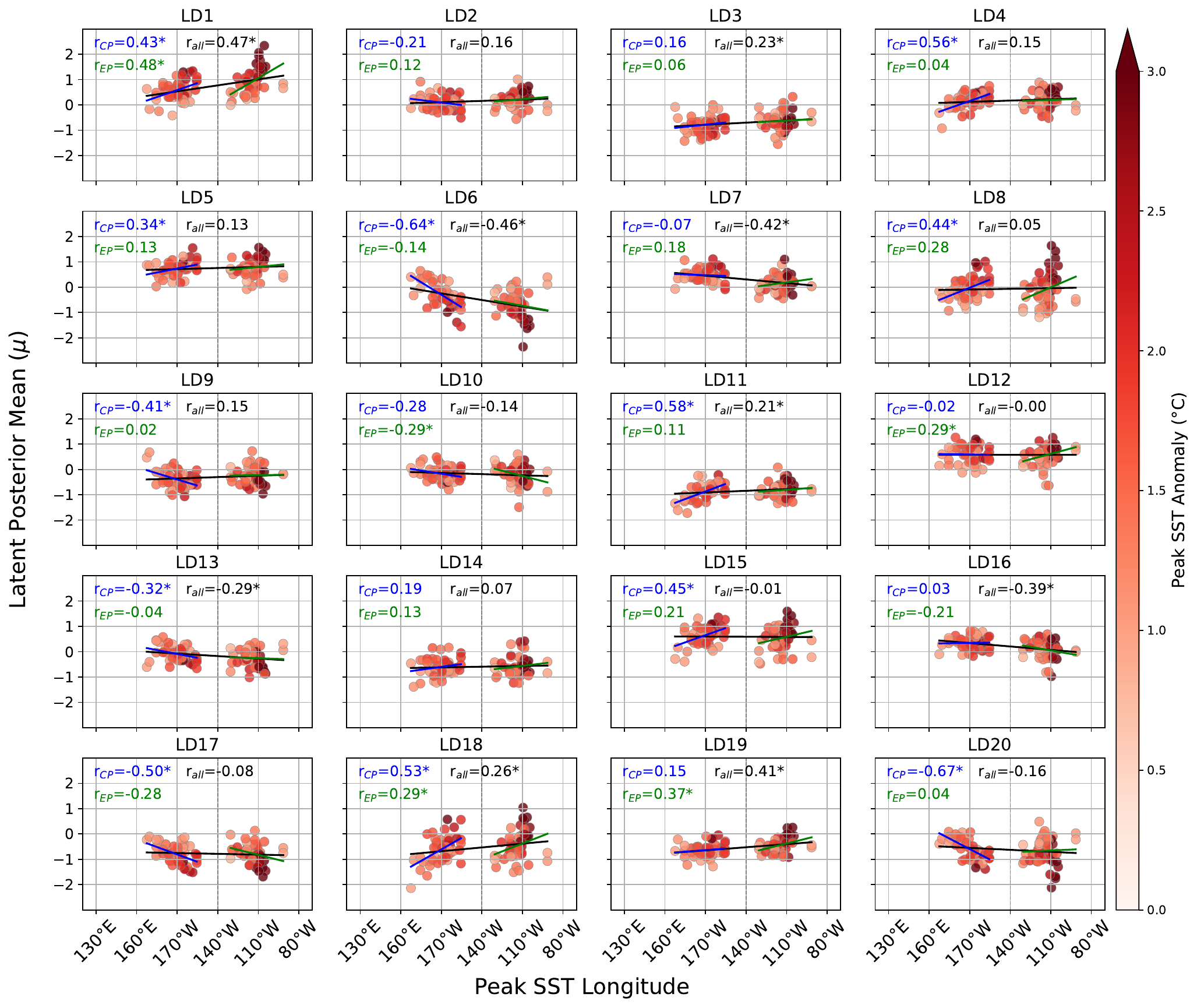}
    \caption{Peak SST longitude and latent posterior mean ($\mu_\phi$) stratified by latent dimension (LD), for ONI-defined El Niño events ($N=92$). Marker color indicates the peak SST anomaly ($^{\circ}C$). Lines of best fit and corresponding Pearson correlation coefficients are shown for all events (black), central Pacific events west of $140^{\circ}$W (blue), and eastern Pacific events east of $140^{\circ}$W (green). Statistical significance is evaluated using a two-sided $t$-test, with $p<0.05$ denoted by an asterisk.}
    \label{fig:peaklon_nino}
\end{figure*}

For ONI-defined El Niño events (Figure \ref{fig:peaklon_nino}), two clusters of peak SST longitudes are apparent in each latent dimension, with one in the CP ($\approx180^{\circ}$E) and another in the EP ($\approx110^{\circ}$W). LD1, LDs 6-7, LD16, and LD19 show statistically significant overall correlations ($r_{all}$) that linearly discriminate between EP and CP, suggesting that these dimensions encode ENSO diversity-related information. When stratified by CP and EP, LD17 and LD20 exhibit strong CP correlations ($-0.50$ and $-0.67$, respectively), while LD12 and LD19 show strong EP correlations ($+0.29$, $+0.37$). This result suggests that various latent dimensions may have sensitivity to either EP or CP. Together, these results indicate that while some latent dimensions capture ENSO diversity broadly, others specialize in regime-specific variability that is lost by aggregated metrics (i.e., $r_{all}$).

\begin{figure*}[t]
    \centering
    \includegraphics[width=1\linewidth]{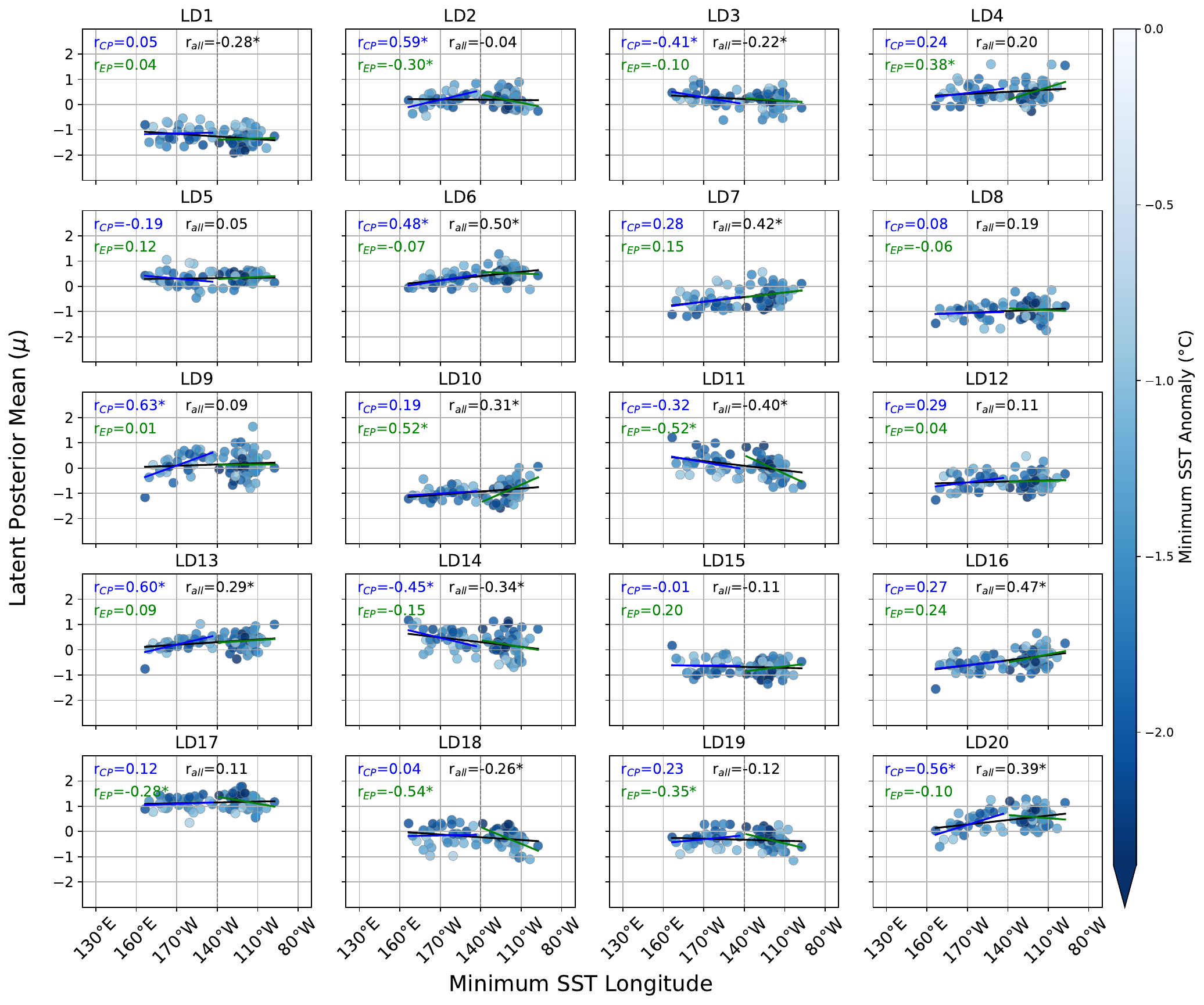}
    \caption{Same as Figure \ref{fig:peaklon_nino}, but using the minimum SST longitude for ONI-defined La Niña events ($N=82$).}
    \label{fig:peaklon_nina}
\end{figure*}

The range of longitudes for peak amplitude SSTs of La Niña events is comparable to that of El Niño events (Figure \ref{fig:peaklon_nina}). However, La Niña events exhibit less spread in latent posterior means than El Niño events. Contrary to the distinct EP and CP clusters seen in Figure \ref{fig:peaklon_nino}, SST peak longitudes for La Niña events are more continuously distributed than bimodal. This interpretation aligns with prior studies suggesting that La Niña diversity is not always cleanly partitioned into EP- and CP-types based on SST patterns alone, and may instead be better characterized as a continuum across the tropical Pacific \citep{ kug2009two, capotondi2015understanding, pan2025diversity}.

LDs 6-7, LD11, LD16, and LD20 exhibit statistically significant overall positive correlations for La Niña events ($r_{all}=+0.39$ to $+0.50$). LD3, LD6, LD9, LDs 13-14, and LD20 correlations are more pronounced in CP than EP La Niña, with $|r_{CP}-r_{EP}|\approx 0.5$. La Niña event EP correlations are generally weak across latent dimensions, with LD10 ($r_{EP}=+0.52$) and LD18 ($r_{EP}=-0.54$) being exceptions and showing strong correlations. Notably, minimum SST anomalies during La Niña events reach approximately $-2^{\circ}C$, which is weaker in magnitude than peak SST anomalies during El Niño events that can reach $+3^{\circ}C$. This asymmetry is consistent with ENSO nonlinearities, wherein La Niña events tend to be weaker than their El Niño counterparts. No single latent dimension linearly discriminates between EP and CP La Niña based on latent posterior means alone.

\begin{figure*}[t]
    \centering
    \includegraphics[width=1\linewidth]{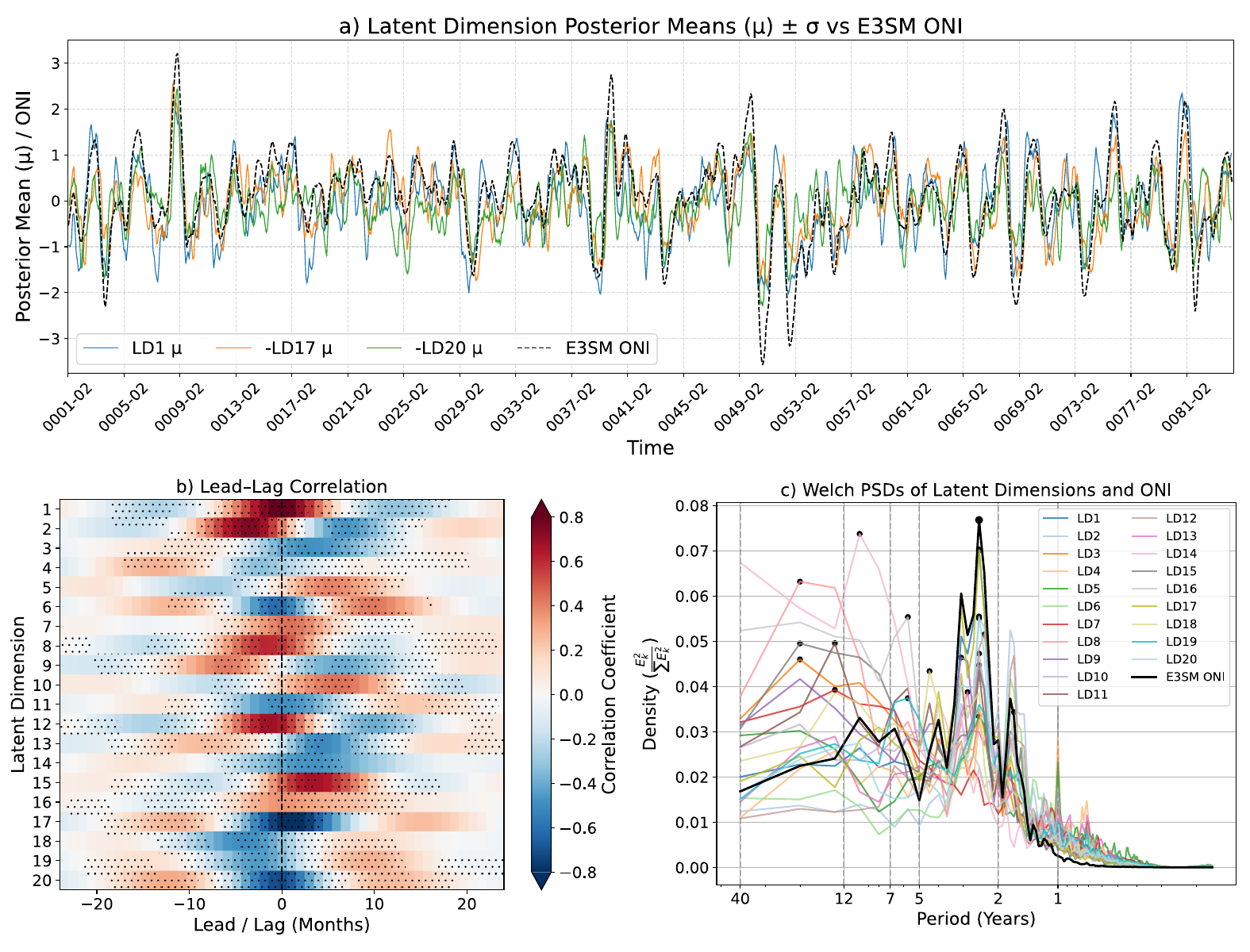}
    \caption{a) Time series of smoothed latent posterior mean ($\mu_{\phi}$, colored lines) and $\pm1$ standard deviation ($\sigma_{\phi}$, shaded lines) for LD1, LD17, and LD20 ($N=5998$). ONI from E3SM is plotted for the corresponding time period (black dashed line). b) Lead-lag correlations between each latent posterior mean (rows) and E3SM ONI are computed over $\pm24$ months (columns). Stippling denotes correlations that are significant at the $99\%$ confidence level based on a two-sided $t$-test that accounts for lag$-1$ (serial) autocorrelation in both time series. c) Power spectral densities (PSDs) of each latent posterior mean and E3SM ONI were computed using Welch's method. The PSDs are normalized to enable direct comparison of total variance across periods. The x-axis is reversed and logarithmic, and the peak period is indicated with a black circle marker on each PSD.}
    \label{fig:multi}
\end{figure*}

We further investigate LD1, LD17, and LD20, which effectively stratified active ENSO phases in Figure \ref{fig:violin}, by examining their temporal evolution alongside ONI derived from E3SM. The latent posterior mean ($\mu_{\phi}$) and standard deviation ($\sigma_{\phi}$) are smoothed using a centered 3-month rolling mean to filter high-frequency variability and mimic ONI temporal smoothing. As a result of the centered window, the first and last months of the smoothed time series were removed ($N=5998$). The resulting spread ($\sigma_{\phi}$) is narrow (Figure \ref{fig:multi}a), consistent with a $\beta$-VAE that emphasizes reconstruction accuracy relative to the KL regularization. However, $\mu_{\phi}$ and $\sigma_{\phi}$ do not have physical units of SST, so they should not be interpreted as SST anomalies. We also computed the zero-lag Pearson correlation between the smoothed (centered 3-month rolling mean) latent posterior means ($N=5998$) and ONI applied to E3SM to quantify agreement between the two time series. The maximum absolute cross-correlation was also computed to identify the strongest linear association across a range of monthly lead/lag values ($N=5998$) over $\pm24$ months, highlighting latent dimensions whose relative phase to ONI is offset rather than concurrent. Correlations in Figure \ref{fig:multi}b and Table \ref{tab:metrics_timeseries} were normalized following the approach outlined in \citet{bretherton1999effective}, and statistical significance was evaluated using a two-sided $t$-test with a sample size that accounts for lag$-1$ autocorrelation in both time series. The `best lag' then corresponds to the time offset (in months) where the absolute correlation is largest. Positive lags indicate that the latent dimension leads ONI, while negative lags indicate that ONI leads the latent dimension.

\begin{table}
\centering
\caption{Metrics comparing E3SM ONI and each latent dimension (LD; $j=1,\dots,d$) smoothed posterior mean ($\mu_{\phi}$; $N=5998$), including Pearson correlation (corr; zero lag), best (absolute maximum) cross correlation and the associated lead/lag ($\pm$months), normalized Dynamic Time Warping (DTW) distances, and the dominant year identified from the Welch power spectral density.}
\begin{tabular}{lccccc}
\toprule
LD & Corr & Best Lag Corr (Month) & Normalized DTW & Dominant Period (Year)\\
\midrule
1 & +0.82 & +0.82 (0) & 0.468 & 2.50 \\
2 & +0.39 & +0.74 (-4) & 0.265 & 2.50 \\
3 & -0.40 & -0.48 (+3) & 0.344 & 20.00 \\
4 & -0.26 & -0.33 (-2) & 0.396 & 1.67 \\
5 & +0.13 & +0.40 (+5) & 0.379 & 2.50 \\
6 & -0.60 & -0.61 (-1) & 0.314 & 2.50 \\
7 & +0.43 & +0.44 (+1) & 0.283 & 13.33 \\
8 & +0.56 & +0.60 (-2) & 0.391 & 20.00 \\
9 & -0.05 & +0.42 (-6) & 0.288 & 3.08 \\
10 & +0.20 & +0.44 (+6) & 0.413 & 2.50 \\
11 & -0.49 & -0.49 (0) & 0.396 & 13.33 \\
12 & +0.64 & +0.68 (-1) & 0.310 & 2.35 \\
13 & -0.17 & -0.51 (+4) & 0.310 &  2.86\\
14 & -0.42 & -0.48 (+3) & 0.385 & 10.00 \\
15 & +0.47 & +0.66 (+3) & 0.284 & 20.00 \\
16 & +0.36 & +0.36 (0) & 0.274 & 5.71 \\
17 & -0.81 & -0.83 (+1) & 0.363 & 2.50 \\
18 & -0.32 & -0.50 (-4) & 0.381 & 4.44 \\
19 & -0.35 & -0.45 (-3) & 0.375 & 5.71 \\
20 & -0.72 & -0.72 (0) & 0.325 & 2.50 \\
\bottomrule
\end{tabular}
\label{tab:metrics_timeseries}
\end{table}

Overall, there is temporal alignment and sign consistency between LD1 and ONI (Figure \ref{fig:multi}a), which is also evident in the (zero-lag) correlation of $+0.82$ (Table \ref{tab:metrics_timeseries}). LD17 and LD20 also exhibit large-magnitude correlations ($-0.81$ and $-0.72$, respectively), which we plot as $-$LDs in Figure \ref{fig:multi}a to align with the sign of E3SM ONI. Overall, the shared temporal structure between the latent dimensions and ONI reinforces captured ENSO variability, albeit not perfectly. LD1, LD17, and LD20 exhibit a small negative bias relative to ONI ($-0.14$, $-0.11$, and $-0.21$). Departures from ONI in amplitude and timing suggest that the latent dimensions may encode ENSO variability not captured by the SST-based ONI. Differences in amplitude are expected because the latent posterior means correspond to compressed representations of SST, OHC, and OLR. Although not shown in Figure \ref{fig:multi}a, LD11 and LD16 exhibit moderate correlations with ONI ($-0.49$ and $+0.36$, respectively; Table \ref{tab:metrics_timeseries}). LD5, LD9, and LD13 exhibit the weakest (zero-lag) correlations with ONI E3SM ($+0.13$, $-0.05$, and $-0.17$).

Several latent dimensions exhibit coherent lead-lag structure with E3SM ONI, indicating that the $\beta$-VAE latent space captures multiple phases of ENSO evolution rather than only contemporaneous variability (Figure \ref{fig:multi}b). At positive leads, LD5, LD7, LD10, and LD15 exhibit sustained positive correlations over lead times of $1-12$ months, whereas LD3, LDs 13-14, and LD17 show sustained negative correlations over similar lead intervals. These patterns suggest that some latent dimensions contain precursor-like variability that leads ONI, either with the same or opposite sign. At negative leads, LD2 and LD12, and more weakly LD8 and LD9, show positive correlations over lags of $1-10$ months, indicating variability that follows ONI. The sign reversal in LD2 and LD12, with negative correlations at positive leads and positive correlations at negative leads, suggests that these latent dimensions encoded a temporally shifted component of ENSO-related variability, potentially associated with preconditioning during event development and a different relationship during the mature or decay phase. Weaker negative lagged correlations are also present in LD6, LD18, and LD19 over adjacent lags of about $-1$ to $-4$ months. Many lead-lag bands are statistically significant (black stippling; Figure \ref{fig:multi}b), supporting the interpretation that the temporal relationships are systematic rather than isolated features. The correlations that peak beyond $\pm12$ months further suggest that these relationships are strongest on decadal or longer timescales (Table \ref{tab:metrics_timeseries}).

We use Dynamic Time Warping \cite[DTW;][]{sakoe2003dynamic} to quantify similarity in temporal evolution between ONI and the smoothed latent posterior means ($N=5998$). In contrast to correlation, DTW computes a distance between two time series by allowing one series to be nonlinearly stretched or compressed along the time axis, thereby aligning similar features, even if they occur at slightly different times. We compute normalized DTW as the sum of absolute differences along the optimal warping path in the ONI–latent-dimension pairwise distance matrix, divided by the number of matched steps, such that lower values indicate greater similarity.

LD2, LD7, LD9, and LDs 15-16 have relatively low normalized DTW ($<0.30$), suggesting that they reproduce ONI-like evolution after allowing flexible time alignment (Table \ref{tab:metrics_timeseries}). This result holds even when their correlations are only modest or inverted in sign, indicating that these latent dimensions can preserve ENSO-like temporal structure without being strong pointwise replicas of ONI. In contrast, LD1 and LD10 have higher normalized DTW values ($>0.40$), indicating that strong correlation does not necessarily imply close agreement in their full temporal evolution. The absence of a monotonic relationship between Pearson correlation and DTW therefore suggests that the latent space separates different aspects of ENSO-related variability, including amplitude covariance, lagged behavior, and similarities in event timing and progression.

To examine the dominant temporal variability of the smoothed latent posterior means and E3SM ONI ($N=5998$), we compute the power spectral density (PSD) using the Welch method (Figure \ref{fig:multi}c). We select the Welch method because its segmentation-and-averaging approach yields a more stable estimate of the PSD than a single periodogram \citep{welch1967use}. The latent posterior mean and E3SM ONI were divided into $40$-year segments with $50\%$ overlap, with the segment mean removed and a Hann window applied prior to computing each periodogram. This was done to provide a smoothed estimate of spectral power while retaining the ability to resolve interannual-to-decadal variability. No detrending was applied, as the E3SM piControl contains no long-term forced trend. The resulting PSDs were then normalized so that the area under each curve sums to 1, enabling direct comparison of the relative contributions of different periods to the total variance. Periods exceeding $40$ years are excluded because they are poorly sampled within the windowed segments. 

We find that $13$ of $20$ latent posterior means exhibit their largest spectral peak between $2$ and $7$ years, corresponding with the canonical ENSO band. E3SM ONI peaks at $2.50$ years, matching the dominant period of LDs 1-2, LDs 5-6, LD10, LD17, and LD20. Although many latent dimensions share this dominant ENSO timescale, earlier analyses (Figure \ref{fig:multi}a,b) indicate that they are associated with distinct expressions of ENSO-related variability, including differences in amplitude and sign. On the other hand, $6$ LDs peak at periods longer than $7$ years. These lower-frequency peaks suggest that the learned representations also capture variability outside the canonical ENSO band, spanning decadal timescales. Plausible candidates include Pacific decadal variability, such as the Pacific Decadal Oscillation (PDO) and tropical Pacific decadal variability (TPDV), whose spatial structure and temporal evolution overlap with or are influenced by ENSO \citep{kirtman1998decadal,rodgers2004tropical}. Previous literature has suggested the PDO and TPDV modulate the background state of the tropical Pacific on decadal timescales, influencing ENSO amplitude, frequency, and spatial pattern \citep{choi2013enso,sun2020impact,power2021decadal}. This may explain why some $\beta$-VAE LDs encode these modes of decadal variability alongside ENSO-related variability.

\begin{figure*}[t]
    \centering
    \includegraphics[width=1\linewidth]{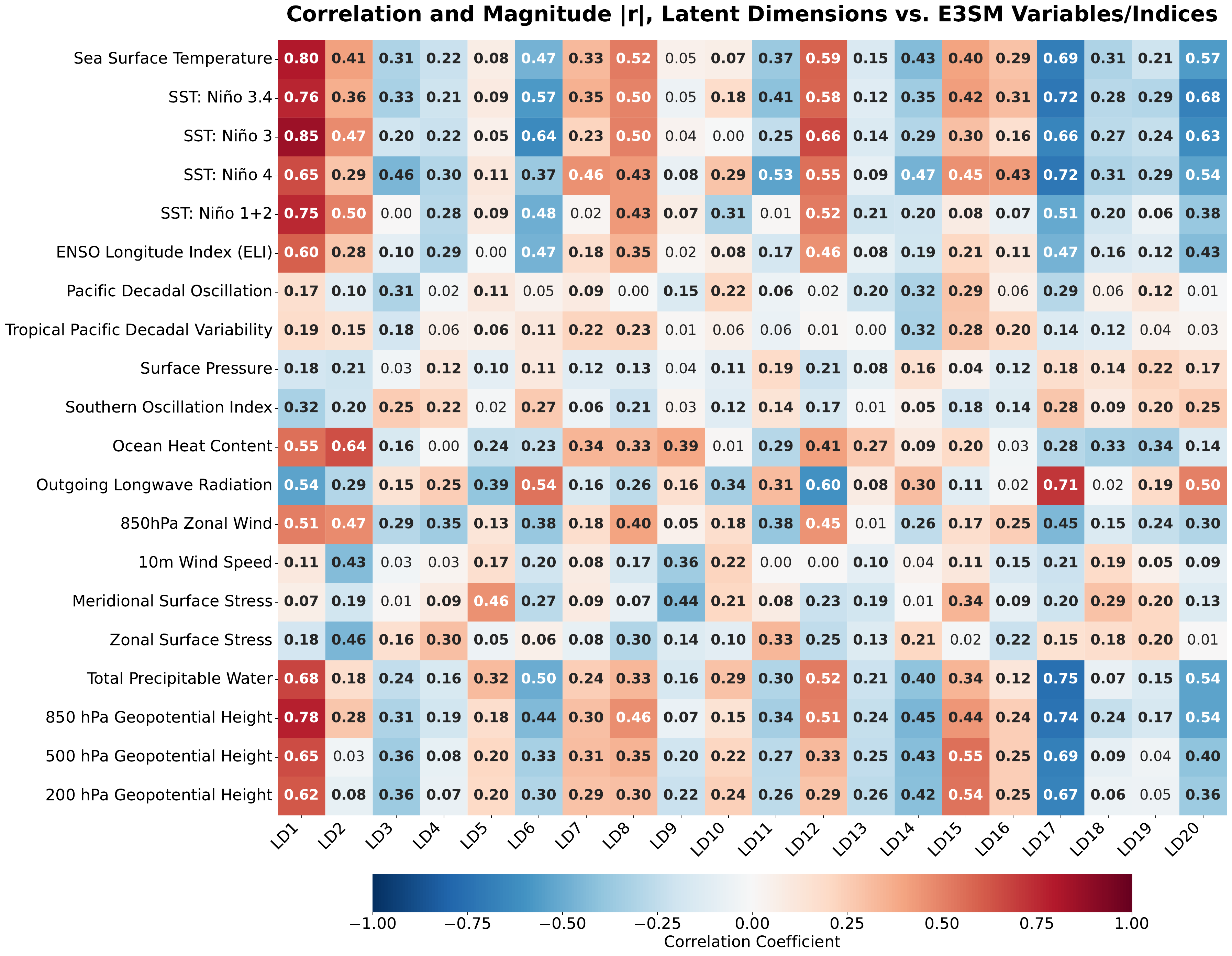}
    \caption{Pearson correlation coefficient matrix between (unsmoothed) latent posterior means and variables from E3SM piControl ($N=6000$). Boldface text indicates correlations that are statistically significant at the $\alpha = 0.01$ threshold based on a two-sided t-test.}
    \label{fig:corr}
\end{figure*}

To quantify the relationship between the latent space and a broad set of climatological variables related to tropical Pacific variability, we compute Pearson correlations between each (unsmoothed) latent posterior mean ($N=6000$) and variables from the E3SMv2 piControl (Figure \ref{fig:corr}). Variables were spatially averaged over the $\beta$-VAE spatial domain and include the original input fields (OLR, SST, OHC), surface pressure, geopotential height ($850$, $500$, and $200$ hPa), zonal and meridional surface stress, $850$ hPa zonal wind, $10$m wind speed, and total precipitable water. We also compute correlations with various ENSO SST regions applied to E3SM (e.g., Niño 1+2), which are defined over their standard regions rather than the entire $\beta$-VAE domain. The following indices were also derived: ENSO Longitude Index (ELI), the Tropical Pacific Decadal Variability (TPDV), and the Pacific Decadal Oscillation (PDO) (further details are provided in the Appendix). These regions and variables are used to interpret the $\beta$-VAE latent space, rather than to evaluate performance.

LD1, LD12, LD17, and LD20 exhibit moderate-to-strong correlations with ENSO indices, geopotential height, OLR, total precipitable water, and 850 hPa zonal wind, indicating that these dimensions capture coupled ENSO-like variability across the ocean and atmosphere. Correlations with geopotential heights from 850 to 200 hPa are often stronger than those with surface pressure or SOI, suggesting that the $\beta$-VAE preferentially captures large-scale circulation patterns rather than surface-based atmospheric fields, which is expected; no near-surface atmospheric fields are used as inputs into the $\beta$-VAE. LD3, LD7, LD11, LDs 14-16, and LD19 show systematically stronger correlations in Niño 4 than Niño 1+2, consistent with the westward-displaced SST anomaly centers. LD3, LD7, and LD11 correlations are near zero with Niño 1+2, but remain moderate with Niño 4 ($-0.46$, $+0.46$, and $-0.53$), reinforcing the interpretation that these latent dimensions preferentially encode CP rather than EP variability. LD10 is particularly distinct, with opposing signs of Niño 1+2 and Niño 4 correlations that correspond to the zonal SST dipole (Figure \ref{fig:corr}), further underscoring the spatial diversity represented in the $\beta$-VAE latent space. 

LD4, LD11, and LDs 15-17 exhibit OHC correlations that are weaker than, or opposite in sign to, their SST correlations, consistent with variability expressed more strongly at the surface. In contrast, LD9, LD13, and LD19 show modest OHC correlations without comparable SST correlations, suggesting partial surface-subsurface decoupling. LD10 shows a pronounced zonal SST contrast but essentially no OHC correlation, consistent with a primarily surface-driven mode. The strongest OHC correlation occurs in LD2 ($0.64$). On longer timescales, the PDO and TPDV show statistically significant correlations across many latent dimensions whose dominant periods in Table \ref{tab:metrics_timeseries} are $\geq10$ years, including LD3, LDs 7-8, and LDs 14-15. Several dimensions correlate more strongly with either TPDV or PDO, suggesting the $\beta$-VAE can disentangle variability within and across spectral frequencies, even beyond the tropical Pacific region (e.g., PDO).

\subsection{Custom Latent Traversal}

We next examine how reconstructed SST, OHC, and OLR respond to perturbations along individual latent dimensions using a latent traversal experiment \citep{kim2018disentangling}. In traditional latent traversals, one latent dimension is varied while the remaining dimensions are held fixed, and the resulting effect on the decoded outputs is assessed. In our case, synthetic 20-dimensional latent vectors were constructed by sampling at evenly spaced values from $-3$ to $+3$ for 1 dimension ($N=6000$; informed by Figure \ref{fig:violin}), while all remaining dimensions were fixed at zero. These latent coordinate values were chosen to limit out-of-distribution extrapolation, while centering the fixed latent dimensions at a common value. Synthetic vectors were then passed directly to the decoder, and reconstructed fields were inverse-transformed to the original physical units ($^{\circ}C; \times10^{8}J\,m^{-2}; W\,m^{-2}$). The resulting SST, OHC, and OLR reconstructions characterize the decoder's response to perturbations along the latent coordinate $z_i$ with all other latent coordinates fixed at zero. We note that fixing the remaining latent coordinates at their encoded mean values, rather than at zero, yields broadly consistent results (not shown).

We use a linear sensitivity diagnostic for latent traversals based on the SST, OHC, and OLR decoded fields. At each grid cell $(x,y)$, we regress the reconstructed anomaly $v(x,y)$ against the latent coordinate $z_i$ across the traversal. The resulting slope provides a first-order estimate of

\begin{equation}
    \frac{\partial v(x,y)}{\partial z_j}, \qquad v \in \text{\{SST, OHC, OLR\}},
\end{equation}

\noindent that is, the change in the reconstructed anomaly at location $(x,y)$ associated with a perturbation in latent dimension $j$. This approach provides an interpretable way to assess how tropical Pacific variability may be encoded, rather than relying on mean composites, although the latter do yield spatial patterns broadly similar to those from the linear sensitivity experiment that are more difficult to interpret given the large number of figures (not shown).

\begin{figure*}[t]
    \centering
    \includegraphics[width=1\linewidth]{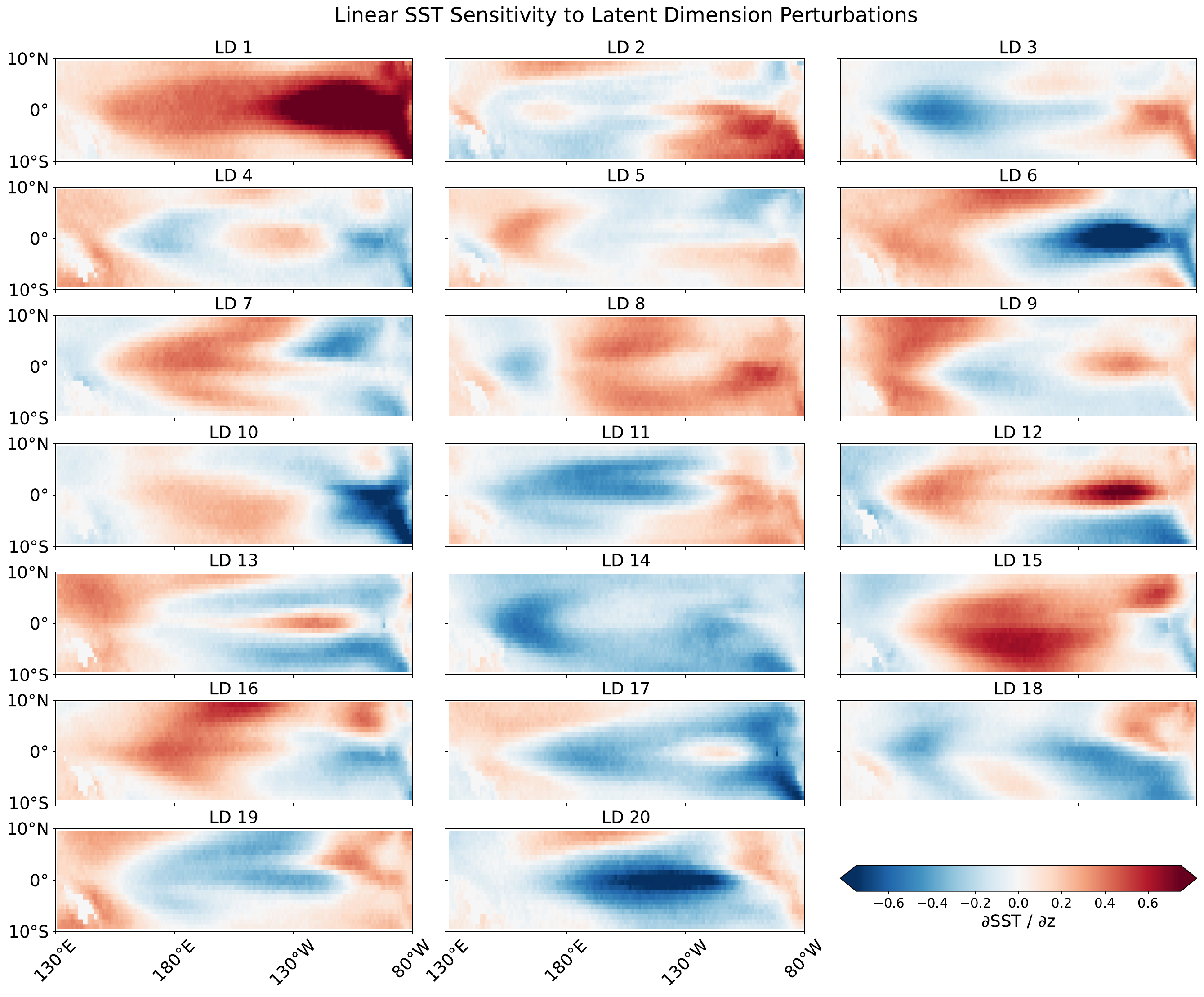}
    \caption{Spatial linear sensitivity of reconstructed SST ($^\circ C$) to perturbations along latent dimensions (LDs), representing the decoder's first-order response to latent perturbations.}
    \label{fig:sst_sensitivity}
\end{figure*}

LD1, LD6, and LD12 show strong EP SST sensitivity, with maximum sensitivity extending westward from the coast of South America and decreasing in amplitude (Figure \ref{fig:sst_sensitivity}), consistent with EP El Niño-like variability. In contrast, LD15 and LDs 19-20 exhibit peak SST sensitivity in the CP. LD7 exhibits a broad region of positive SST sensitivity spanning both the CP and EP. LD2 and LDs 10-11 exhibit enhanced SST sensitivity along the South American coast in the eastern basin, resembling a coastal Niño 1+2-like structure. LDs 3-5, LD9, LD13, and LDs 17-18 exhibit dipole-like structures in their SST sensitivity, with opposing signs across latitude and/or longitude, making interpretation more challenging. LD14 shows negative SST sensitivity across nearly the entire region of interest, suggesting it may not encode SST variability, may reflect a larger-scale mode of variability beyond the tropical Pacific, or may co-vary with another latent dimension. Overall, spatially distinct sensitivity patterns suggest that the $\beta$-VAE latent space may learn to represent and separate different modes of SST variability.

\begin{figure*}[t]
    \centering
    \includegraphics[width=1\linewidth]{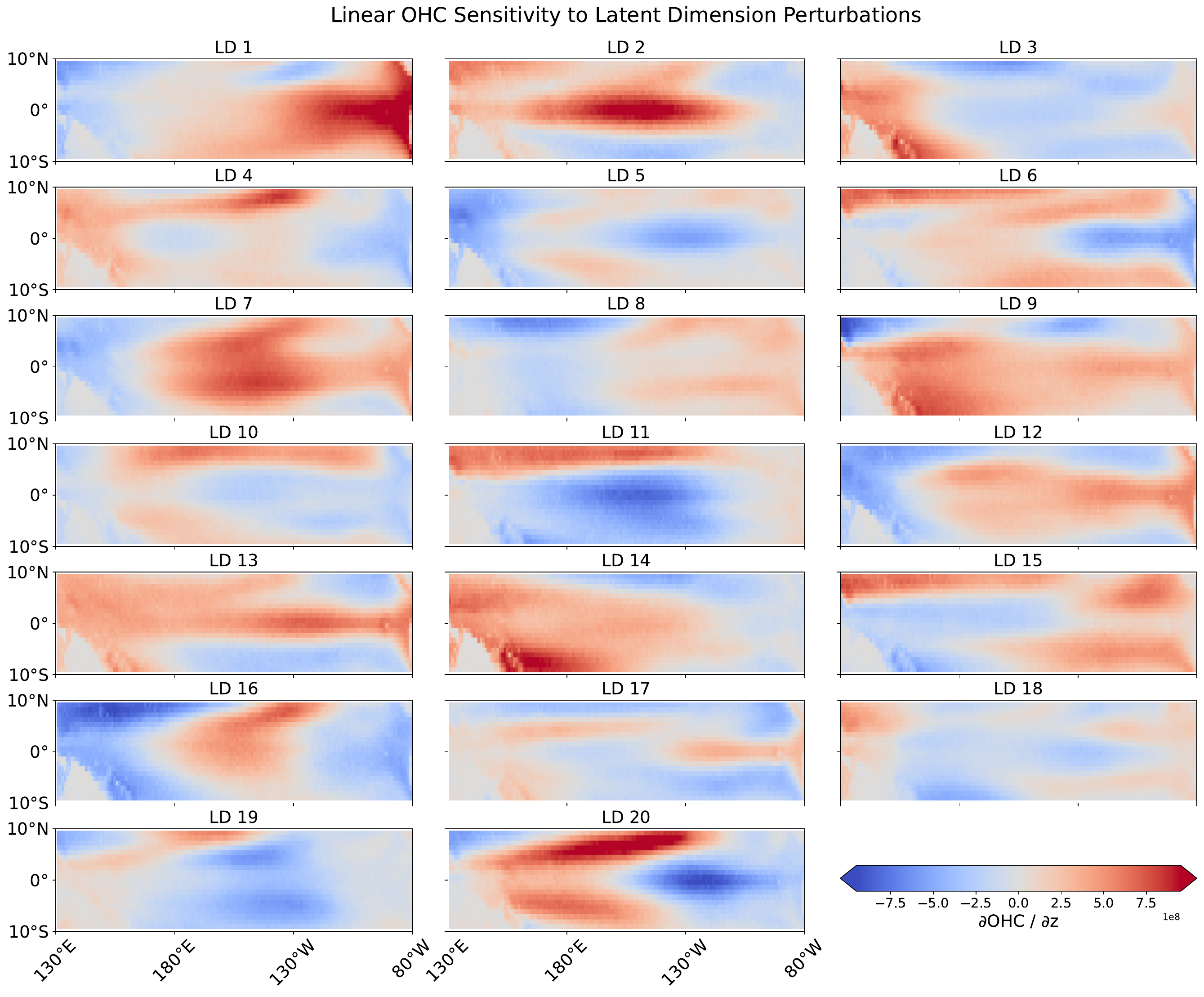}
    \caption{Same as Figure \ref{fig:sst_sensitivity}, but for OHC ($1 \times 10^{8} \ J\,m^{-2}$).}
    \label{fig:ohc_sensitivity}
\end{figure*}

Similar to SST, LD1 exhibits strong EP-like OHC sensitivity, where the subsurface and surface signals appear coupled (Figure \ref{fig:ohc_sensitivity}). Similar but weaker results are evident for LD6 and LD12. CP-like OHC sensitivity is primarily observed in LD2, with weaker expression in LD16 and opposite-signed sensitivity in LD11 and LD19. Since this CP OHC sensitivity is not clearly reflected in the corresponding SST patterns, these latent dimensions may capture relative decoupling between surface and subsurface variability. LD7 and LD12 exhibit both EP and CP-like sensitivity, suggesting overlapping influences. Western Pacific OHC sensitivity is strongest in LD14 and weaker in LD3, while LD9 shows broadly positive basin-wide sensitivity, and LD13 is concentrated in an equatorial band across the CP and EP. Given the weaker and less coherent corresponding SST sensitivities, these longer dominant periods may arise primarily from OHC variability. Off-equatorial sensitivity is also evident in LD4, LD6, LD11, and LD15. In contrast, LD5, LD10, LD17, and LD18 exhibit weaker equatorial OHC sensitivity, suggesting limited representation of canonical ENSO. Overall, these results underscore the heterogeneous distribution of OHC sensitivity across the $\beta$-VAE latent space. 

\newpage

\begin{figure*}[t]
    \centering
    \includegraphics[width=1\linewidth]{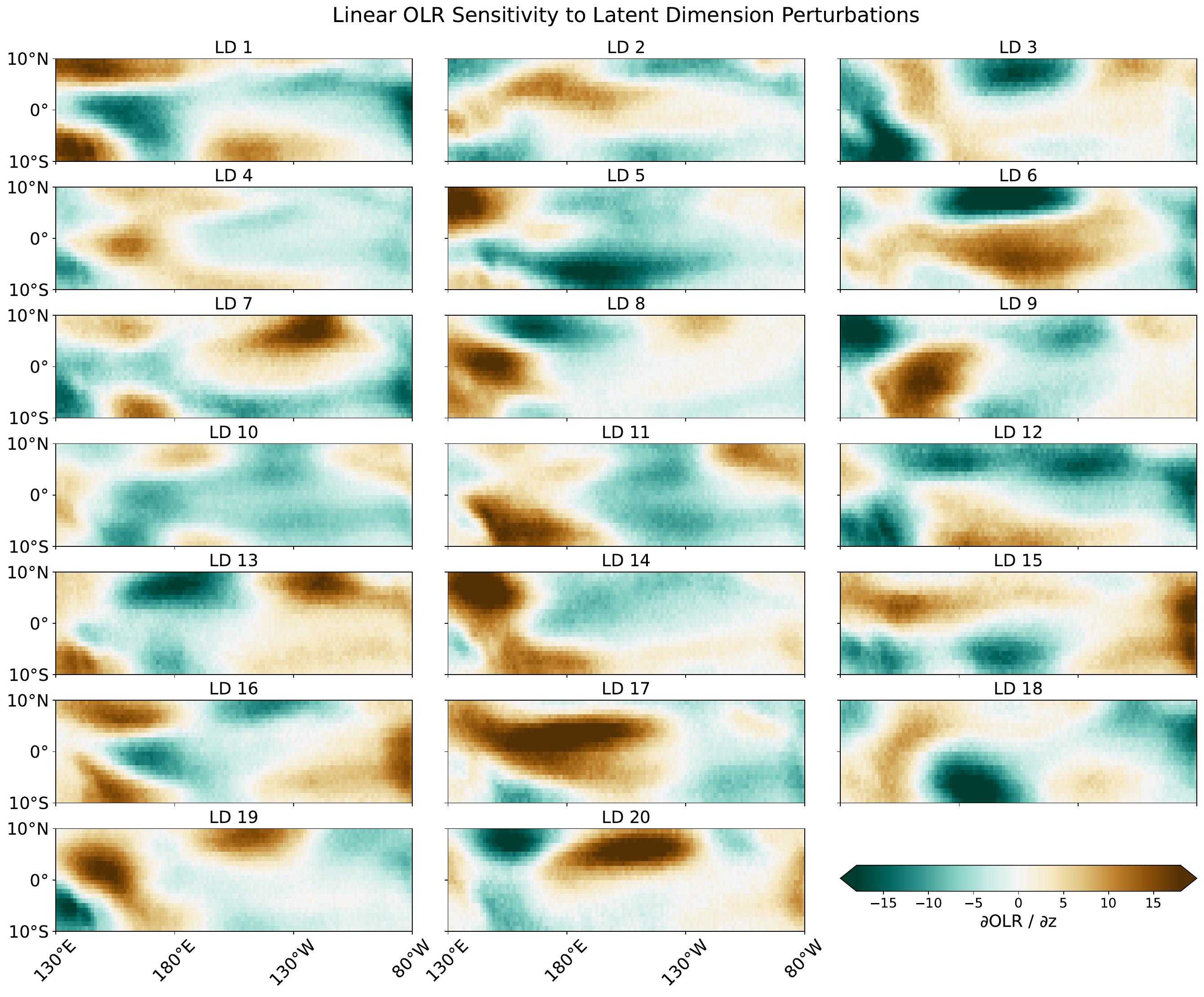}
    \caption{Same as Figure \ref{fig:sst_sensitivity}, but for OLR ($W\,m^{-2}$).}
    \label{fig:olr_sensitivity}
\end{figure*}

The OLR sensitivity analysis shows broader and more spatially heterogeneous responses across latent dimensions than SST or OHC (Figure \ref{fig:olr_sensitivity}). Unlike the more coherent equatorial structures seen in several SST and OHC sensitivity maps, OLR exhibits alternating positive and negative responses across longitude and latitude, often with substantial off-equatorial structure. This result suggests that perturbations to individual latent dimensions tend to redistribute reconstructed OLR spatially rather than produce a single dominant tropical response. As a result, the OLR sensitivity maps are less readily grouped into EP- or CP-like categories than SST or OHC. Instead, they suggest that OLR-related variability is distributed across multiple latent dimensions and expressed through diverse zonal and meridional patterns, consistent with OLR reflecting a combination of convective, cloud, and circulation-related variability. However, sensitivity maps alone do not identify the underlying dynamical mechanisms. Accordingly, these results are best interpreted as showing where reconstructed OLR increases or decreases in response to perturbations in each latent dimension.

\newpage

\section{Conclusions}

We evaluated a multi-branch $\beta$-VAE trained on tropical Pacific SST, OHC, and OLR to assess whether it can skillfully reconstruct coupled climate variability and organize that variability into physically interpretable latent dimensions. The model generalizes well to unseen data, with only modest degradation from training to test performance. Across all three variables, test errors are about $8-10\%$ larger than training errors, while test-set $R^2$ retains about $88-93\%$ of the training-set variance. Spatial reconstruction skill is also consistent with known tropical Pacific variability. SST and OHC are reconstructed more effectively in regions dominated by large-scale coherent oceanic variability, whereas OLR exhibits lower and more spatially heterogeneous skill, consistent with its higher spectral frequency. The contrast between absolute error metrics and $R^2$ also shows that reconstruction performance depends on the local variance structure: regions with relatively large variability can exhibit larger absolute errors but maintain higher explained variance.

\begin{figure*}[t]
    \centering
    \includegraphics[width=1\linewidth]{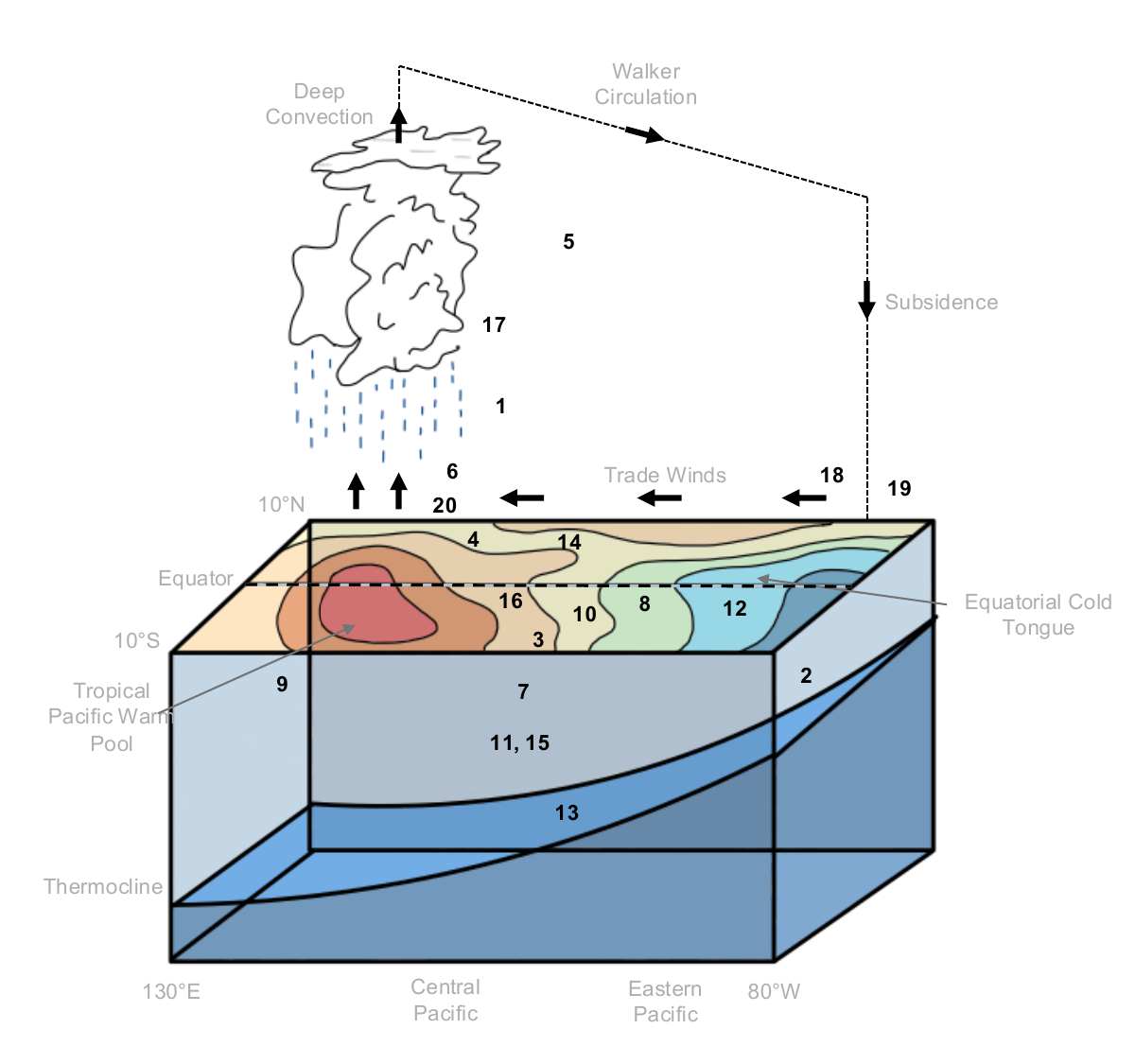}
    \caption{Conceptual schematic summarizing the hypothesized physical interpretation of the $\beta-VAE$ latent dimensions (LDs; $j=1,\dots,d$) in the tropical Pacific. Labels are positioned according to the region, variable, or coupled process with which they are most consistently associated, based on reconstruction diagnostics, statistical relationships, and latent traversal experiments. This schematic provides a qualitative synthesis of subsurface, sea surface, and atmospheric interactions represented by the $\beta$-VAE LDs, rather than uniquely assigning individual LDs to known tropical Pacific processes.}
    \label{fig:schematic}
\end{figure*}

The latent space diagnostics show that the $\beta$-VAE does not organize variability in a PCA-like hierarchy. Instead, the learned representation is variable-dependent and nonlinear. SST-variability is concentrated in a relatively small number of latent dimensions, whereas OHC- and OLR-variability are more broadly distributed. This pattern of encoded variability is consistent with the linear PCA baseline (Section 3\ref{variance_diagnostics}). Several latent dimensions consistently align with known ENSO structure. LD1, LD17, and LD20 most clearly separate ONI-defined El Niño and La Niña conditions, while lead-lag correlations and spectral analyses show that the latent space captures not only concurrent ENSO variability but also temporally shifted and lower-frequency modes of climate variability (e.g., TPDV and PDO). The latent transversal experiments further indicate that different latent dimensions encode distinct spatial responses, including EP-like, CP-like, coastal, and subsurface-dominated structures. Figure \ref{fig:schematic} summarizes our qualitative interpretation of the latent dimensions in the context of coupled tropical Pacific variability, and a dimension-by-dimension summary is provided in Table \ref{tab:ENSOrelation}. LD labels are placed in the schematic based on qualitative analysis of Figures \ref{fig:exp1_vae}-\ref{fig:violin} and \ref{fig:corr}-\ref{fig:olr_sensitivity}, considering where their associated linear sensitivity magnitudes were largest and which variable they were more strongly associated with in terms of variance and statistical correlation (e.g., subsurface, surface, or atmosphere). Several latent dimensions did not yield results readily linked to tropical Pacific variability, suggesting that either the $\beta$-VAE was unable to extract physically interpretable information (e.g., architecture-related) or that the latent dimensions may be linked to processes that are presently unrecognized contributors to variability in the tropical Pacific basin, which could be potentially explored in future work.

Overall, the $\beta$-VAE provides a skillful and physically informative nonlinear representation of coupled tropical Pacific variability. It preserves large-scale SST, OHC, and OLR structures while learning latent dimensions corresponding to distinct aspects of ENSO and related ocean-atmosphere coupling. The $\beta$-VAE may also enable investigation into mode onset and decay for predictability or more targeted teleconnection indices \cite[e.g.,][]{passarella2023assessing}. The generative capabilities of the $\beta$-VAE may also enable the creation of synthetic data to extend limited observational records \cite[e.g.,][]{kadow2020artificial}. However, several limitations also exist. For example, the latent space is only partially disentangled, particularly for atmospheric variability and for distinctions such as EP versus CP ENSO. Future work could systematically evaluate alternative reconstruction loss functions (e.g., MAE, Huber, LogCosh) to analyze their impact on the $\beta$-VAE latent space. We also find that the choice of ablation baseline (mean- vs zero-replacement) may push the decoder outside of the learned latent manifold and amplify latent dimension variability. Future work using similar variance-diagnostic frameworks should account for this sensitivity when selecting and interpreting baseline replacement experiments. A plausible contributor to LDs representing multiple modes of variability is feature superposition, where a network encodes more features than its latent space, leading to overlapping directions \citep{elhage2022toy}. Sparse autoencoders (SAEs) have been proposed to address this through an encoder-decoder model that projects data into a sparse latent space while using an $L1$ penalty to ensure that each active dimension corresponds to a single feature \citep{cunningham2023sparse,king2025leveraging}. Whether the assumption of sparse, independent features translates to representing intrinsically coupled climate-scale processes in the tropical Pacific remains an open question, making SAEs a compelling avenue for future work. A challenge with training a $\beta$-VAE is the need for a sufficiently large training dataset; the perfect model framework used herein helped overcome the limited sample size of observed ENSO events, but in exchange, tropical Pacific biases from E3SM were likely inherited by the $\beta$-VAE \cite[e.g.,][]{fasullo2023overview,fasullo2024modes}. Future work should explore the utility of transfer learning with reanalyses to mitigate model biases and nonstationary extrapolation, although recent work has shown that a sufficiently long reanalysis record is required for such applications \citep{mayer2025can}. Nevertheless, the $\beta$-VAE shows promise for reduced-dimensional analyses of nonlinear climate variability and for identifying interpretable latent structure in the coupled Earth system.

\clearpage
\acknowledgments
This material is based upon work supported by the US National Science Foundation (NSF) Graduate Research Fellowship Program under Grant No. DGE 2236417. Any opinions, findings, and conclusions or recommendations expressed in this material are those of the author(s) and do not necessarily reflect the views of the US NSF. This material is also based upon work supported by the U.S. DOE, Office of Science, BER, RGMA component of the Earth and Environmental System Modeling Program under Award \#DE-SC0024093. The Computational and Information Systems Laboratory at the US NSF National Center for Atmospheric Research (NCAR), a major facility sponsored by the US NSF under Cooperative Agreement No. 1852977, provided computing and data storage resources. The use of Grammarly (\url{https://app.grammarly.com/}) and ChatGPT (\url{https://chat.openai.com/}) is acknowledged to refine academic language and improve the flow of the writing. All AI-generated content has been reviewed and edited to ensure accuracy, and full responsibility is taken for the final content of this manuscript.

%
%
\datastatement
The data for this study are available on the Earth System Grid (e.g., \url{https://aims2.llnl.gov/search/?project=E3SM/}, E3SM Project, 2024). Github:

\clearpage
\FloatBarrier
\appendix
\appendixtitle{$\beta$-VAE Training and Interpretation Materials}\label{appendref}

\section{Training Loss Monitoring}

\begin{figure}[t]
    \centering
    \includegraphics[width=0.85\linewidth]{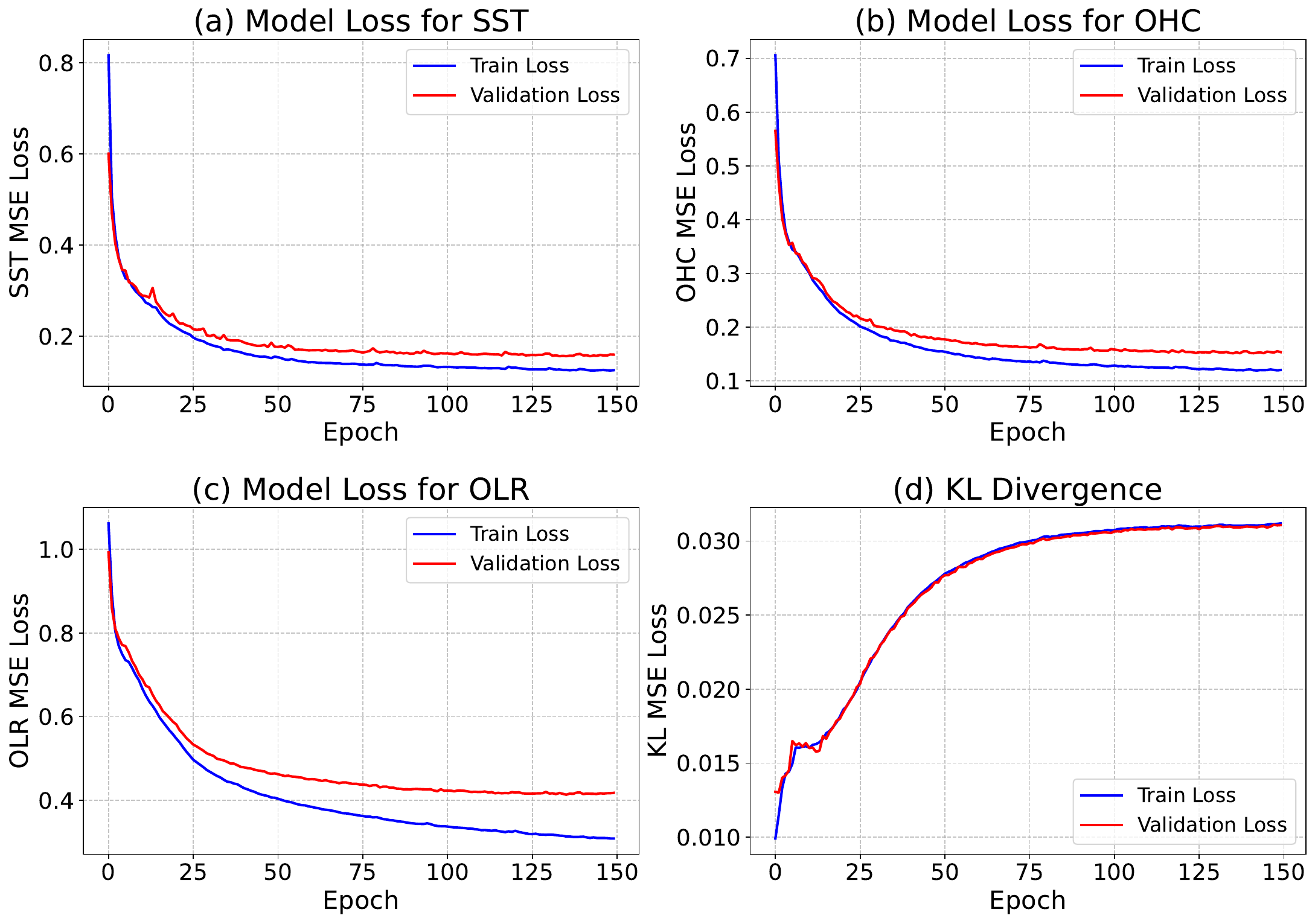}
    \caption{Total loss as a function of epoch for the training (blue) and validation sets (red) for the multivariate $\beta$-VAE. MSE loss is shown for a) SST, b) OHC, and c) OLR. d) Shows the KL divergence term.}
    \label{fig:vae_loss}
\end{figure}

The four terms comprising the custom loss function were monitored independently during training and validation, and are available in Figure \ref{fig:vae_loss}.

\section{$\beta$-VAE Dimensionality Sensitivity}
\label{sec:dimensionality}

To assess how latent dimensionality affects the $\beta$-VAE representation, models with $d \in \{5,10,15,20,25,30\}$ latent dimensions were trained using $\beta=0.0005$. Each configuration was assessed using two complementary metrics: the dimension-normalized dispersion of latent variance ratios $\mathrm{LVR}_{dispersion}$ and the variance across latent dimensions of the spatial reconstruction sensitivity obtained from the normalized latent-traversal slope maps $\mathrm{Var(spatial\_var)}$.

For each latent dimension $j$,

\begin{equation}
\mathrm{LVR}_j = \frac{\mathrm{Var}(z_j)}{\sum_{k=1}^{d} \mathrm{Var}(z_k)}, \qquad j=1,\dots,d,
\end{equation}
where $d$ is the total number of latent dimensions. By construction,

\begin{equation}
\sum_{j=1}^{d}\mathrm{LVR}_j=1 \qquad \text{and} \qquad \overline{\mathrm{LVR}}=\frac{1}{d}.
\end{equation}
The population variance across latent dimensions is then defined as
\begin{equation}
    \mathrm{Var}\left(\left\{\mathrm{LVR}_j\right\}_{j=1}^{d}\right)=\frac{1}{d}\sum_{j=1}^{d}\left(\mathrm{LVR}_j-\frac{1}{d}\right)^2.
\end{equation}
To compare models with different $d$, we normalize by the maximum possible variance, $d^2/(d-1)$, 
\begin{equation}
    \mathrm{LVR_{dispersion}}=\frac{d^2}{d-1}\mathrm{Var}\left(\left\{\mathrm{LVR}_j\right\}_{j=1}^{d}\right).
\end{equation}

\noindent Low $\mathrm{LVR_{dispersion}}$ indicates that latent variance is distributed relatively evenly rather than concentrated in a small subset of latent dimensions and is therefore preferred.

To quantify the spatial sensitivity of the reconstructed fields to each latent dimension, each latent dimension $j$ is traversed independently. At each grid cell $(x,y)$, the reconstructed anomaly for variable $v$ is regressed against the latent coordinate $z_{i,j}$ across traversal samples $i$, producing a slope map, $S_j^v(x,y)$, for each reconstructed field (SST, OHC, and OLR). Each slope map is normalized by the standard deviation of all slope values for that variable $v$ ($\sigma_v$). The spatial sensitivity associated with latent dimension $j$, $\mathrm{spatial\_var}_j$, is then defined as the average spatial variance of its normalized slope maps across the three reconstructed variables $V=3$:

\begin{equation}
\mathrm{spatial\_var}_j = \frac{1}{V}\sum_{v=1}^V \mathrm{Var}_{x,y}\left(\frac{S_j^v(x,y)}{\sigma_v}\right), \qquad V=3.
\end{equation}

\noindent $\mathrm{Var(spatial\_var)}$ then follows the form:

\begin{equation}
\mathrm{Var(spatial\_var)} = \mathrm{Var}\left(
\left\{\mathrm{spatial\_var}_j\right\}_{j=1}^{d}
\right), \qquad j=1,\dots,d.
\end{equation}

\noindent High $\mathrm{Var(spatial\_var})$ indicates that latent dimensions differ substantially in the magnitude of the spatial sensitivity they induce in the reconstructed fields, and is preferred.

At $d=5$, $\mathrm{LVR_{dispersion}}$ is large, suggesting that latent variance is concentrated in a small subset of dimensions. As $d$ increases, this concentration generally decreases. In contrast, the dispersion of per-dimension spatial reconstruction sensitivity varies non-monotonically with latent-space dimensionality. At $d=25$ and $d=30$, the differentiation of spatial sensitivity among latent dimensions declines, while $\mathrm{LVR_{dispersion}}$ improves only marginally relative to $d=20$.

\begin{figure}[t]
    \centering
    \includegraphics[width=0.85\linewidth]{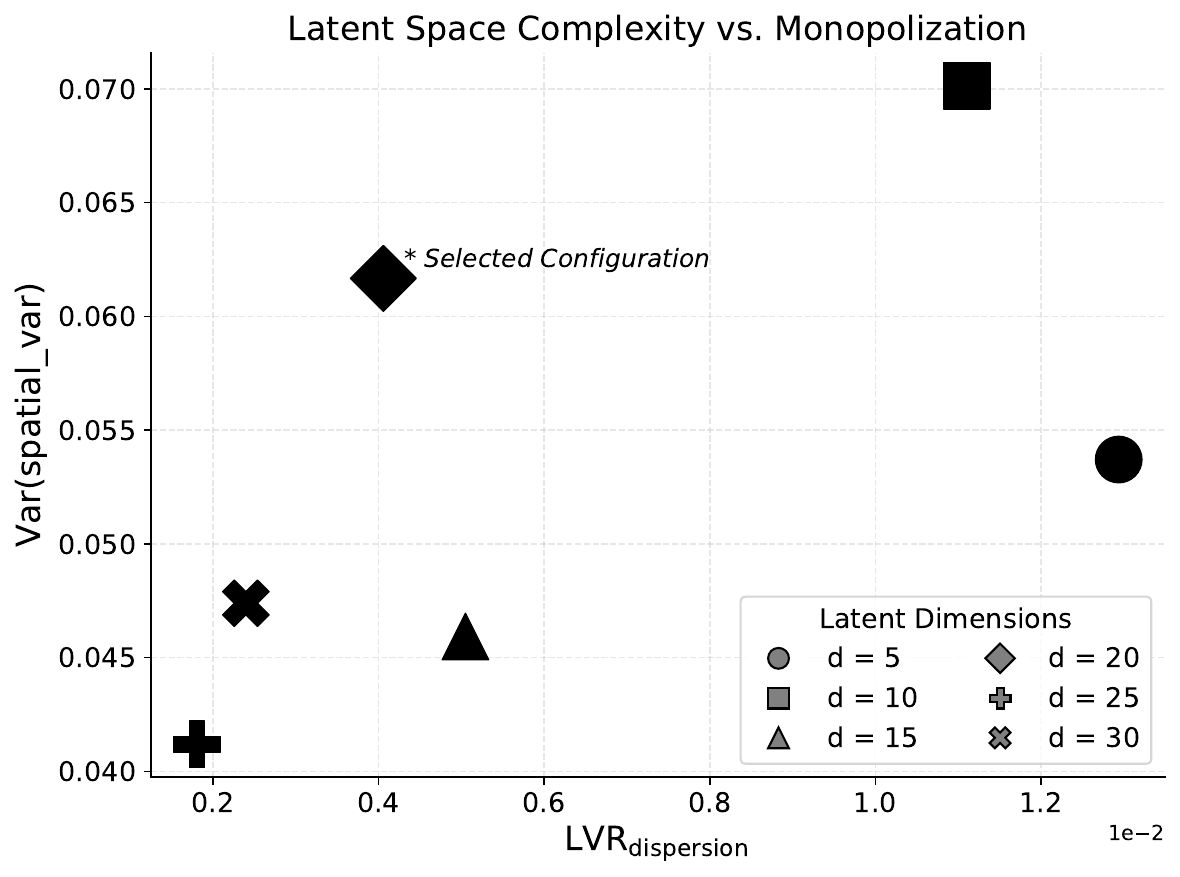}
    \caption{Trade-off between normalized latent-variance dispersion
    ($\mathrm{LVR}_{\mathrm{dispersion}}$; x-axis) and the dispersion of spatial reconstruction sensitivity across latent dimensions ($\mathrm{Var(spatial\_var)}$; y-axis) for $\beta$-VAE configurations with latent-space dimensionality $d \in \{5,10,15,20,25,30\}$, indicated by marker shape. All configurations use $\beta=0.0005$.}
    \label{fig:complexity_tradeoff}
\end{figure}

The preferred configuration in Figure \ref{fig:complexity_tradeoff} is $d=20$, which combines low $\mathrm{LVR_{dispersion}}$ and high $\mathrm{Var(spatial\_var})$. This configuration limits the concentration of variance in a small subset of latent dimensions while maintaining differentiated, dimension-specific levels of spatial reconstruction sensitivity.

\section{$\beta$-VAE Ensembling}
\label{sec:ensemble}

To more rigorously assess whether the $\beta$-VAE is sensitive to random weight initialization, a 50-member ensemble is trained with varying random seeds. Table \ref{tab:ensemble} reports the mean and standard deviation of RMSE and MAE for each input variable in the test set. Standard deviations are less than $1\%$ of the corresponding mean for all variables and metrics considered. Reconstruction error is highly consistent, indicating that model reconstruction skill is not sensitive to weight initialization.

\begin{table}[htbp]
\centering
\caption{Reconstruction test error (RMSE and MAE) for SST, OHC, and OLR across the 50-member ensemble. Mean and standard deviation are reported.}
\begin{tabular}{l c c}
\textbf{Variable} & \textbf{RMSE} & \textbf{MAE} \\
\hline
SST & 0.398 $\pm$ 0.004 & 0.305 $\pm$ 0.003 \\
OHC & 0.390 $\pm$ 0.004 & 0.296 $\pm$ 0.003  \\
OLR & 0.632 $\pm$ 0.004 & 0.460 $\pm$ 0.003 \\
\end{tabular}
\label{tab:ensemble}
\end{table}

\section{$\beta$-VAE Latent Space Reproducibility}
\label{sec:reproducibility}

We then examine the reproducibility of the learned latent representations across ensemble members on the test set ($N=1200$). VAEs impose no constraints on the ordering or sign of latent dimensions across training runs. To account for latent permutations and sign reversals across training runs, a $20\times20$ absolute-correlation matrix is constructed for each ensemble member. Each matrix entry contains the absolute Pearson correlation between one latent dimension from the reference run and one latent dimension from the ensemble member. The reference run is the model used throughout the main text ($\mathrm{seed}=1$) and is trained independently of the ensemble. Optimal permutation matching is then used to identify the one-to-one assignment that maximizes the summed absolute correlation across the 20 matched LD pairs. 

\begin{figure}[t]
    \centering
    \includegraphics[width=1\linewidth]{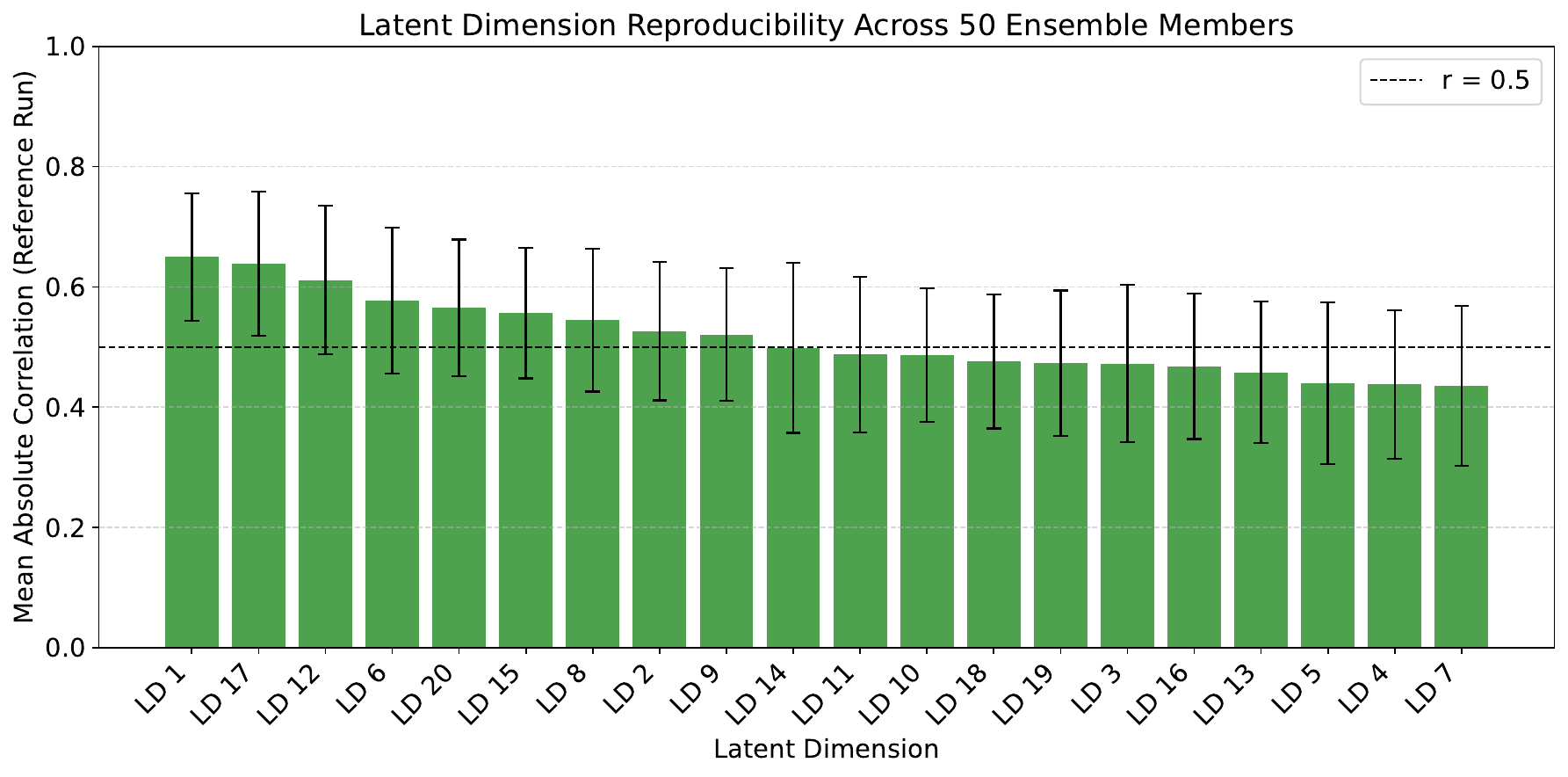}
    \caption{Mean absolute correlation $\overline{|r|}$ of each latent dimension with the corresponding reference run ($\mathrm{seed}=1$) after optimal permutation alignment across 50 ensemble members. Error bars represent $\pm 1$ standard deviation across ensemble members. The horizontal dashed line (black) shows $\overline{|r|}=0.5$. Dimensions are sorted in descending order of mean absolute correlation.}
    \label{fig:latent_reproducibility}
\end{figure}

Figure \ref{fig:latent_reproducibility} shows the mean absolute Pearson correlation across ensemble members for each reference latent dimension after optimal permutation matching. The most reproducible latent dimensions are LD1 and LD17 ($\overline{|r|}=0.65, 0.64$), whereas the least reproducible latent dimensions are LD4 and LD7 ($\overline{|r|}=0.44, 0.44$). LD1 and LD17 also exhibit strong correlations with ENSO-related signals in previous analyses (e.g., Figures \ref{fig:multi}-\ref{fig:corr} and Table \ref{tab:metrics_timeseries}), suggesting that their greater reproducibility may reflect their representation of dominant ENSO-related variability. Across all LDs, mean absolute correlations are $\overline{|r|} \sim 0.5$, indicating partial but incomplete correspondence among training runs.

\section{$\beta$-VAE Ensemble Variance Experiments}
\label{sec:reproducibility}

\begin{figure}[t]
    \centering
    \includegraphics[width=1\linewidth]{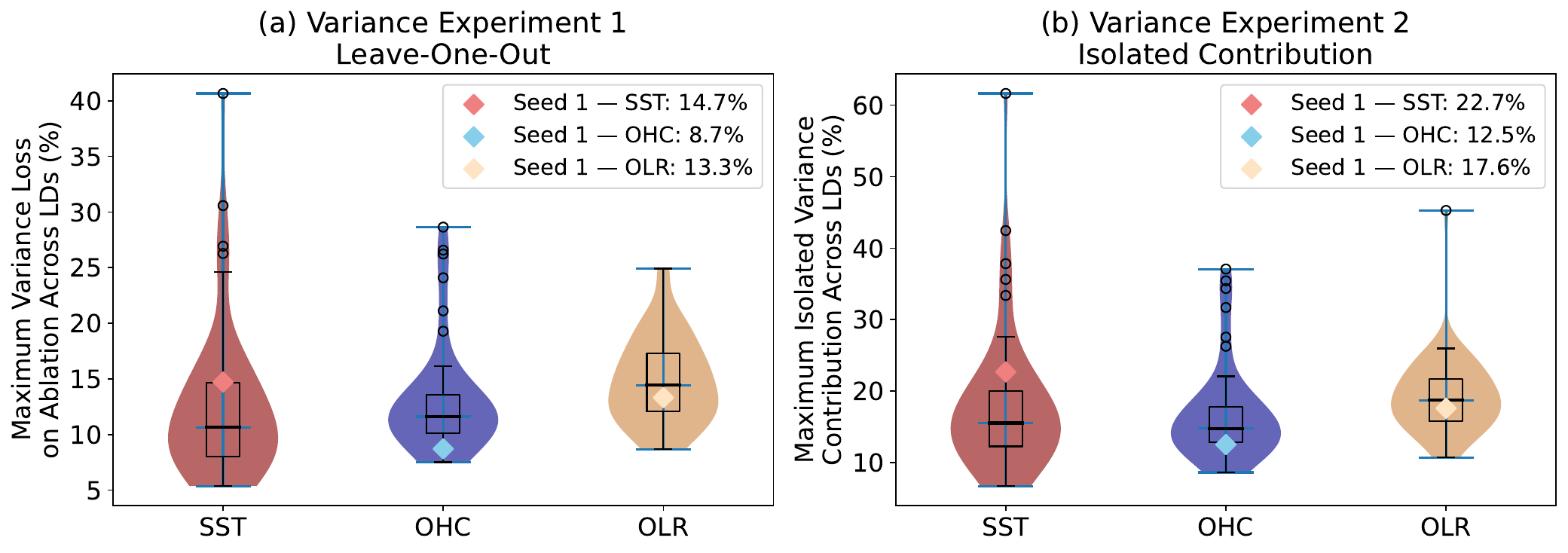}
    \caption{Distribution of latent dimension variance loss and contribution across the 50-member $\beta$-VAE ensemble for SST (red), OHC (blue), and OLR (orange). Panel (a) represents the maximum variance ablation loss across all latent dimensions, corresponding to Experiment 1. Panel (b) shows the maximum isolated variance generation across all latent dimensions, corresponding to Experiment 2. Violin shapes indicate the full distribution across seeds, boxes represent the interquartile range and median, whiskers extend to $1.5 \times IQR$, and open circles denote outlier ensemble members. Diamonds denote the reference run ($\mathrm{seed}=1$).}
    \label{fig:ensemble_ablation}
\end{figure}

To assess whether the magnitude of the strongest latent-dimension contribution is reproducible across ensemble members (experiments as in Figures \ref{fig:exp1_vae}-\ref{fig:violin}), we evaluate the maximum variance contribution separately for SST, OHC, and OLR. For each ensemble member and variable, we calculate two complementary metrics across all latent dimensions: (1) the variance loss induced by ablating each latent dimension using that ensemble member's corresponding encoder-decoder pair, and (2) the variance generated when each latent dimension is varied in isolation. For each metric, we retain the maximum percentage contribution across latent dimensions. The latent dimension producing this maximum may differ among ensemble members, so this analysis does not require permutation matching or alignment. These metrics follow the variance-diagnostic frameworks described in Section 3\ref{variance_diagnostics}. Since VAEs do not enforce consistent latent-dimension ordering across initializations, averaging LD-specific contributions according to their optimal permutation matching may combine dimensions that are not equivalent across ensemble members. Computing LD ensemble means to reproduce Figures \ref{fig:exp1_vae} and \ref{fig:exp2_vae} with 50 ensemble members may therefore dampen contributions across dimensions. Nevertheless, Figures S1 and S2, which reproduce Figures \ref{fig:exp1_vae} and \ref{fig:exp2_vae} with 50 ensemble members, are included in the supplemental material for reference. To avoid averaging contributions from latent dimensions that may not be aligned across ensemble members, we instead show the violin distributions of the maximum variance loss or contribution across the 50-member ensemble for each variable (Figure \ref{fig:ensemble_ablation}). These distributions characterize the magnitude, but not the identity, of the dominant latent-dimension contribution in each ensemble member. For Experiment 1, the mean dominant LD ablation loss across seeds is $12.6 \pm 6.8\%$ for SST, $13.0 \pm 4.8\%$ for OHC, and $14.9 \pm 3.8\%$ for OLR. For Experiment 2, the mean isolated variance contribution is $18.1 \pm 9.8\%$ for SST, $16.7 \pm 6.5\%$ for OHC, and $19.4 \pm 5.3\%$ for OLR. 

The $\beta$-VAE 50-member ensemble variance experiments and the reference run (seed=1) variance experiments are evaluated on the same held-out test set (N=1200), allowing direct comparison of the reference run’s dominant contributions against the ensemble distribution. For both Experiments 1 and 2, the reference run’s dominant OLR contribution (13.3\%, 17.6\%) falls within the ensemble’s interquartile range. In both Experiments 1 and 2, the SST contribution lies above the ensemble’s interquartile range (14.7\%, 22.7\%), while the OHC contribution falls slightly below (8.7\%, 12.5\%). The reference run’s dominant contributions, evaluated on the held-out test set, are therefore broadly consistent with the ensemble distribution. The distributions nevertheless include ensemble members with substantially larger contributions, including an Experiment 1 SST ablation loss of 40.7\%, indicating that the concentration of variance in the dominant latent dimension can vary appreciably across model initializations.

\section{Reconstruction Evaluation}

MAE and RMSE were computed at each reconstructed grid cell using the following equations,

\begin{equation}
    \text{MAE} = \frac{1}{T}\sum_{t=1}^{T} \big|\hat{X}_t - X_t\big|,
\end{equation}

\begin{equation}
    \text{RMSE}= \sqrt{\frac{1}{T}\sum_{t=1}^{T}(\hat{X}_t - X_t)^2},
\end{equation}

\noindent where $t$ is the time step, $T$ is the number of samples ($T = 1200$ for the test set and $T = 4800$ for the training set), $X_t$ is the input value, and $\hat{X}_t$ is the reconstructed input value. To compute $R^2$, we first calculated the temporal variance of the input data time series $X=\{X_t\}_{t=1}^{T}$ at each grid cell,

\begin{equation}
    \mathrm{Var}(X) = \frac{1}{T}\sum_{t=1}^{T}(X_t - \bar{X})^2,
\end{equation}

\noindent where

\begin{equation}
    \bar{X}=\frac{1}{T}\sum_{t=1}^{T}X_t.
\end{equation}

\noindent Grid cells with zero temporal variance ($\mathrm{Var}(X)=0$) are excluded. For grid cells where temporal variance is not zero, we computed the mean squared error (MSE):

\begin{equation}
    \mathrm{MSE} = \frac{1}{T}\sum_{t=1}^{T}(\hat{X}_t - X_t)^2,
\end{equation}

\noindent and then $R^2$ as

\begin{equation}
    R^2 = 1 - \frac{\mathrm{MSE}}{\mathrm{Var}(X)}.
\end{equation}

\begin{table}
\centering
\caption{Reconstruction skill of the $\beta$-VAE for SST ($^{\circ}C$), OHC ($\times 10^{8}J m^{-2}$), and OLR ($Wm^{-2}$) on the 1200-sample test set, with 4800-sample training set results for comparison. RMSE, MAE, and $R^2$ are computed per sample over valid grid points. Reported values are sample means, with standard deviations (STD) in parentheses, along with skill ratios (SRs; test/train). Uncertainty is estimated by nonparametric bootstrapping ($B=1000$) with resampling over time steps. Two-tailed 95\% confidence intervals (CI) are given for the test minus train performance difference.}
\begin{tabular}{lcccccc}
\toprule
Metric & Variable & Test (STD) & Train (STD) & SR & CI (95\%) \\
\midrule
MAE & SST & 0.229 (0.041) & 0.211 (0.033) & 1.087 & [0.016, 0.021] \\
MAE & OHC & 2.705 (0.497) & 2.455 (0.417) & 1.102 & [0.219, 0.280]  \\
MAE & OLR & 6.604 (1.573) & 6.003 (1.388) & 1.100 & [0.499, 0.688]  \\
\midrule
RMSE & SST & 0.296 (0.052) & 0.273 (0.043) & 1.083 & [0.020, 0.026] \\
RMSE & OHC & 3.504 (0.625) & 3.201 (0.529) & 1.095 & [0.262, 0.341]\\
RMSE & OLR & 8.896 (1.866) & 8.133 (1.654) & 1.094 & [0.640, 0.867] \\
\midrule
$R^2$ & SST & 0.594 (0.260) & 0.653 (0.216) & 0.910 & [-0.075, -0.042] \\
$R^2$ & OHC & 0.661 (0.221) & 0.713 (0.183) & 0.926 & [-0.066, -0.040] \\
$R^2$ & OLR & 0.539 (0.199) & 0.612 (0.163) & 0.881 & [-0.085, -0.060] \\
\bottomrule
\end{tabular}
\label{tab:metrics}
\end{table}

Table \ref{tab:metrics} reports MAE, RMSE, and $R^2$ with standard deviations, skill ratios, and $95\%$ confidence intervals for test minus train sets across all three variables. Table \ref{tab:metrics} shows that while test set MAE and RMSE are consistently higher than training MAE and RMSE for all variables, the magnitude of this degradation is small relative to the variability in per-sample errors (indicated by the standard deviations, i.e., STD). RMSE exceeds MAE across all variables, likely due to larger reconstruction errors from outliers. However, RMSE remains within the same order of magnitude as MAE.

\section{E3SM ELI Calculation}

We computed the ENSO Longitude Index (ELI) following \citet{williams2018diversity} using E3SMv2 piControl monthly SSTs over the equatorial Pacific ($5^{\circ}$S$-5^{\circ}$N, $120^{\circ}$E$-80^{\circ}$W). For each month, the convective threshold was defined as the mean SST over $30^{\circ}$S$-30^{\circ}$N, and equatorial Pacific grid points with SSTs at or above this threshold were identified as convectively favorable. ELI was then computed as the mean longitude of those points, yielding a monthly time series of the zonal position of deep convection.

\section{E3SM TPDV Calculation}

The TPDV index was computed from monthly SSTAs, with monthly climatology removed, over the tropical Pacific domain ($20^{\circ}$S$-20^{\circ}$N, $120^{\circ}$E$-80^{\circ}$W), following \citet{power2021decadal}. ENSO was removed at each grid via linear regression of the Niño 3.4 index ($5^{\circ}$S$-5^{\circ}$N, $190^{\circ}$E$-120^{\circ}$W) onto the SSTAs at each grid cell and subtracting the fitted component, analogous to \citet{zhang1997enso}. The residual field was then smoothed with a $72$-month centered rolling mean. The latitude-weighted spatial mean was then taken, yielding a time series.

\section{E3SM PDO Calculation}

We computed the PDO index following \citet{mantua1997pacific} using monthly SSTs from the E3SMv2 piControl over the North Pacific ($20^{\circ}$N$-70^{\circ}$N, $110^{\circ}$E$-100^{\circ}$W). The monthly mean climatology was removed at each grid point, and a linear trend was fit and subtracted along the time dimension before taking the latitude-weighted mean of the SSTAs. The PDO index was defined as the standardized principal component of the leading EOF of the area-weighted SST anomaly field. The sign convention was adjusted so that the positive phase corresponds to anomalously cool SSTs in the central North Pacific and warm SSTs along the North American coast \citep{newman2016pacific}.

\section{Latent Space Physical Interpretation}

The corresponding Table \ref{tab:ENSOrelation} for Figure \ref{fig:schematic} is provided herein.

\begin{table}[htbp]
\centering
\caption{Qualitative interpretations of the 20 $\beta$-VAE latent dimensions (LDs; $j=1,\dots,d$) based on latent space diagnostics, statistical analyses, and latent traversal experiments. Interpretations summarize the subsurface, sea surface, and atmospheric interactions represented by each latent dimension.}
\begin{tabular}{c c}
\textbf{LD} & \textbf{Description} \\
\hline
1 & EP-like ENSO variability with strongly coupled SST, OHC, and OLR. \\
2 & Coastal Niño-like SST variability with pronounced CP subsurface expression. \\
3 & CP-like SST variability, most evident in the Niño 4 region. \\
4 & Weak SST-variability with modest coupled atmosphere-ocean structure. \\
5 & Predominantly atmospheric variability. \\
6 & Strong coupled SST-atmosphere variability. \\
7 & CP-like variability (Niño 4) with both surface and subsurface ocean expression. \\
8 & Mixed EP- and CP-like SST variability with coupled SST, OHC, and OLR structure. \\
9 & Primarily subsurface-dominated variability with modest surface expression. \\
10 & Zonal contrast between EP- and CP-like SST variability (Niño 1+2 and Niño 4). \\
11 & CP-like variability with moderate subsurface and atmospheric expression. \\
12 & EP-like ENSO variability with strongly coupled SST, OHC, and OLR structure. \\
13 & Predominantly subsurface-dominated variability. \\
14 & CP-like variability with weak subsurface and moderate atmospheric expression. \\
15 & CP-like variability with enhanced OHC and moderate atmospheric expression. \\
16 & CP-like variability with weak subsurface coupling. \\
17 & Atmosphere-dominant variability with strong SST-atmosphere coupling. \\
18 & Weak surface-subsurface coupling with variable atmosphere contribution. \\
19 & Weak surface-subsurface coupling with modest large-scale structure. \\
20 & Strong SST-atmosphere coupling with weak subsurface expression. \\
\end{tabular}\label{tab:ENSOrelation}
\end{table}

\clearpage
\bibliographystyle{ametsocV6}
\bibliography{references}

\end{document}